\documentclass{osa-article}
\usepackage{booktabs}
\usepackage{datetime}
\usepackage{endnotes}
\usepackage{hyperref}
  \def\Snospace~{\S{}\,}

\usepackage{makecell}
\usepackage{multirow}
\usepackage{siunitx}
\usepackage{subfig}
\usepackage{threeparttable}

\journal{osajournal}

\articletype{Research Article}

\begin{document}

\title{Broadband, millimeter-wave antireflection coatings for large-format, cryogenic aluminum oxide optics}

\author{
A.~Nadolski,\authormark{1} 
J.~D.~Vieira,\authormark{1,2} 
J.~A.~Sobrin,\authormark{3,4} 
A.~M.~Kofman,\authormark{1,5} 
P.~A.~R.~Ade,\authormark{6} 
Z.~Ahmed,\authormark{7,8} 
A.~J.~Anderson,\authormark{3,9} 
J.~S.~Avva,\authormark{10} 
R.~Basu Thakur,\authormark{3} 
A.~N.~Bender,\authormark{3,11} 
B.~A.~Benson,\authormark{3,9,12} 
L.~Bryant,\authormark{13} 
J.~E.~Carlstrom,\authormark{3,4,11,12,13} 
F.~W.~Carter,\authormark{3,11} 
T.~W.~Cecil,\authormark{11} 
C.~L.~Chang,\authormark{3,11,12} 
J.~R.~Cheshire IV,\authormark{1,14} 
G.~E.~Chesmore,\authormark{15} 
J.~F.~Cliche,\authormark{16} 
A.~Cukierman,\authormark{10} 
T.~de~Haan,\authormark{10} 
M.~Dierickx,\authormark{17} 
J.~Ding,\authormark{18} 
D.~Dutcher,\authormark{3,4} 
W.~Everett,\authormark{19} 
J.~Farwick,\authormark{1} 
K.~R.~Ferguson,\authormark{20} 
L.~Florez,\authormark{1} 
A.~Foster,\authormark{21} 
J.~Fu,\authormark{1} 
J.~Gallicchio,\authormark{3,22} 
A.~E.~Gambrel,\authormark{3} 
R.~W.~Gardner,\authormark{13} 
J.~C.~Groh,\authormark{10} 
S.~Guns,\authormark{10} 
R.~Guyser,\authormark{1} 
N.~W.~Halverson,\authormark{19,23} 
A.~H.~Harke-Hosemann,\authormark{1,24,25} 
N.~L.~Harrington,\authormark{10} 
R.~J.~Harris,\authormark{1,26} 
J.~W.~Henning,\authormark{3,11} 
W.~L.~Holzapfel,\authormark{10} 
D.~Howe,\authormark{27} 
N.~Huang,\authormark{10} 
K.~D.~Irwin,\authormark{7,8,28} 
O.~Jeong,\authormark{10} 
M.~Jonas,\authormark{9} 
A.~Jones,\authormark{27} 
M.~Korman,\authormark{21} 
J.~Kovac,\authormark{17} 
D.~L.~Kubik,\authormark{9} 
S.~Kuhlmann,\authormark{11} 
C.-L.~Kuo,\authormark{7,8,28} 
A.~T.~Lee,\authormark{10,29} 
A.~E.~Lowitz,\authormark{3} 
J.~McMahon,\authormark{15} 
J.~Meier,\authormark{1} 
S.~S.~Meyer,\authormark{3,4,12,13} 
D.~Michalik,\authormark{27} 
J.~Montgomery,\authormark{16} 
T.~Natoli,\authormark{3,12} 
H.~Nguyen,\authormark{9} 
G.~I.~Noble,\authormark{16} 
V.~Novosad,\authormark{18} 
S.~Padin,\authormark{3} 
Z.~Pan,\authormark{3,4} 
P.~Paschos,\authormark{13} 
J.~Pearson,\authormark{18} 
C.~M.~Posada,\authormark{18} 
W.~Quan,\authormark{3,4} 
A.~Rahlin,\authormark{3,9} 
D.~Riebel,\authormark{27} 
J.~E.~Ruhl,\authormark{21} 
J.T.~Sayre,\authormark{19} 
E.~Shirokoff,\authormark{3,12} 
G.~Smecher,\authormark{31} 
A.~A.~Stark,\authormark{17} 
J.~Stephen,\authormark{13} 
K.~T.~Story,\authormark{7,28} 
A.~Suzuki,\authormark{29} 
C.~Tandoi,\authormark{1} 
K.~L.~Thompson,\authormark{7,8,28} 
C.~Tucker,\authormark{6} 
K.~Vanderlinde,\authormark{30,32} 
G.~Wang,\authormark{11} 
N.~Whitehorn,\authormark{20} 
V.~Yefremenko,\authormark{11} 
K.~W.~Yoon\authormark{7,8,28} 
 and M.~R.~Young\authormark{32}
}
\address{
\authormark{1}Department of Astronomy, University of Illinois Urbana-Champaign, 1002 West Green Street, Urbana, IL, 61801, USA\\
\authormark{2}Department of Physics, University of Illinois Urbana-Champaign, 1110 West Green Street, Urbana, IL, 61801, USA\\
\authormark{3}Kavli Institute for Cosmological Physics, University of Chicago, 5640 South Ellis Avenue, Chicago, IL, 60637, USA\\
\authormark{4}Department of Physics, University of Chicago, 5640 South Ellis Avenue, Chicago, IL, 60637, USA\\
\authormark{5}Department of Physics and Astronomy, University of Pennsylvania, 209 South 33rd Street, Philadelphia, PA, 19104, USA\\
\authormark{6}School of Physics and Astronomy, Cardiff University, Cardiff CF24 3YB, United Kingdom\\
\authormark{7}Kavli Institute for Particle Astrophysics and Cosmology, Stanford University, 452 Lomita Mall, Stanford, CA, 94305, USA\\
\authormark{8}SLAC National Accelerator Laboratory, 2575 Sand Hill Road, Menlo Park, CA, 94025, USA\\
\authormark{9}Fermi National Accelerator Laboratory, MS209, P.O. Box 500, Batavia, IL, 60510, USA\\
\authormark{10}Department of Physics, University of California, Berkeley, CA, 94720, USA\\
\authormark{11}High-Energy Physics Division, Argonne National Laboratory, 9700 South Cass Avenue., Argonne, IL, 60439, USA\\
\authormark{12}Department of Astronomy and Astrophysics, University of Chicago, 5640 South Ellis Avenue, Chicago, IL, 60637, USA\\
\authormark{13}Enrico Fermi Institute, University of Chicago, 5640 South Ellis Avenue, Chicago, IL, 60637, USA\\
\authormark{14}Minnesota Institute for Astrophysics, University of Minnesota, 115 Union Street SE, Minneapolis, MN, 55455, USA\\
\authormark{15}Department of Physics, University of Michigan, 450 Church Street, Ann Arbor, MI, 48109, USA\\
\authormark{16}Department of Physics and McGill Space Institute, McGill University, 3600 Rue University, Montreal, Quebec H3A 2T8, Canada\\
\authormark{17}Center for Astrophysics | Harvard \& Smithsonian, 60 Garden Street, Cambridge, MA, 02138, USA\\
\authormark{18}Materials Sciences Division, Argonne National Laboratory, 9700 South Cass Avenue, Argonne, IL, 60439, USA\\
\authormark{19}CASA, Department of Astrophysical and Planetary Sciences, University of Colorado, Boulder, CO, 80309, USA \\
\authormark{20}Department of Physics and Astronomy, University of California, Los Angeles, CA, 90095, USA\\
\authormark{21}Department of Physics, Center for Education and Research in Cosmology and Astrophysics, Case Western Reserve University, Cleveland, OH, 44106, USA\\
\authormark{22}Harvey Mudd College, 301 Platt Boulevard., Claremont, CA, 91711, USA\\
\authormark{23}Department of Physics, University of Colorado, Boulder, CO, 80309, USA\\
\authormark{24}Department of Astrophysical and Planetary Sciences, University of Colorado - Boulder, 2000 Colorado Avenue, Boulder, CO, 80305, USA\\
\authormark{25}National Institute for Standards and Technology, 325 Broadway, Boulder, CO, 80305, USA\\
\authormark{26}Verizon Media Group, 1908 South First Street, Champaign, IL, 61820, USA\\
\authormark{27}University of Chicago, 5640 South Ellis Avenue, Chicago, IL, 60637, USA\\
\authormark{28}Deptartment of Physics, Stanford University, 382 Via Pueblo Mall, Stanford, CA, 94305, USA\\
\authormark{29}Physics Division, Lawrence Berkeley National Laboratory, Berkeley, CA, 94720, USA\\
\authormark{30}Dunlap Institute for Astronomy \& Astrophysics, University of Toronto, 50 St. George Street, Toronto, ON, M5S 3H4, Canada\\
\authormark{31}Three-Speed Logic, Inc., Vancouver, B.C., V6A 2J8, Canada\\
\authormark{32}Department of Astronomy \& Astrophysics, University of Toronto, 50 St. George Street, Toronto, ON, M5S 3H4, Canada\\
}

\email{\authormark{*}nadolsk1@illinois.edu}

\begin{abstract}
We present two prescriptions for broadband ($\sim$77--252 GHz), millimeter-wave antireflection coatings for cryogenic, sintered polycrystalline aluminum oxide optics: one for large-format (\SI{700}{\milli\meter} diameter) planar and plano-convex elements, the other for densely packed arrays of quasi-optical elements, in our case \SI{5}{\milli\meter} diameter half-spheres (called ``lenslets'').
The coatings comprise three layers of commercially-available, polytetrafluoroethylene-based, dielectric sheet material.
The lenslet coating is molded to fit the \SI{150}{\milli\meter} diameter arrays directly while the large-diameter lenses are coated using a tiled approach.
We review the fabrication processes for both prescriptions then discuss laboratory measurements of their transmittance and reflectance.
In addition, we present the inferred refractive indices and loss tangents for the coating materials and the aluminum oxide substrate.
We find that at \SI{150}{\giga\hertz} and \SI{300}{\kelvin} the large-format coating sample achieves \SI[separate-uncertainty=true]{97\pm2}{\percent} transmittance and the lenslet coating sample achieves \SI[separate-uncertainty=true]{94\pm3}{\percent} transmittance.
\end{abstract}

\section{Introduction}
\label{sec:intro}
Current cosmic microwave background (CMB) experiments employ large diameter, efficient refractive optics \cite{thornton:2016,inoue:2016,hui:2018,galitzki:2018,spt3ginstrumentpaper}.
Such optics will also be critical for next-generation CMB experiments, like CMB-S4 \cite{abazajian:2016,abitbol:2017}.
Large, strong, low dielectric loss refractive optics can be constructed from silicon and sintered polycrystalline aluminum oxide (alumina).
Materials such as silicon and alumina allow for optical designs with thinner, lower-curvature lenses than those employing lenses made from low-index plastic such as high-density polyethylene (HDPE) or ultra-high-molecular-weight polyethylene (UHMWPE) \cite{yoon:2008,kermish:2012}.
Additionally, alumina and silicon lenses offer thermal and mechanical advantages: their relatively high thermal conductivity allows them to be cooled to cryogenic temperatures (\SI{\sim 4}{\kelvin}) with small thermal gradients across the optic; their rigid structures and low coefficients of thermal expansion (CTE) allow the lenses to resist deformation.

High purity, single-crystal silicon has a favorably high refractive index and low dielectric loss, but the maximum size of a silicon lens is limited by the size of commercially available silicon boules (\SI{< 45}{\centi\meter}).
These limits are mainly driven by pressure from the semiconductor industry, which uses the boules for batch device fabrication \cite{shimura:2017}.
Alumina optics offer performance similar to silicon in important regions of the CMB observation spectrum, but they can be made with \SI{> 70}{\centi\meter} diameter.
The size of a single alumina part is limited by the diameter of the isostatic press used to compact the raw powder.
However, if alumina parts larger than the press diameter are required, then sections can be compacted separately and fused into a single body at the sintering step \cite{cavanaugh:2020}.
In refracting telescope designs, the lens diameter limits the instrument's resolution, so it is desirable to use as large a lens as possible.
The lens diameter is especially important for low-frequency observations (e.g. within the \SIrange{\sim 30}{\sim 40}{\giga\hertz} atmospheric transmission window).
In addition, large-diameter optics allow for a focal plane with a large useable area, which provides more space for detectors.
For experiments with detectors that are approximately photon shot-noise-limited, increasing the number of detectors is a means of increasing the experiment's sensitivity.
The South Pole Telescope's third-generation survey camera (SPT-3G) employs such large-format, cryogenic, alumina refractive optics.

The high refractive indices ($n\sim3$) of alumina and silicon optics necessitate antireflection (AR) coatings to reduce reflections and maximize the overall instrument sensitivity.
Developing cryogenic AR coatings for high-index optical elements is therefore an area of significant interest to the CMB instrumentation community---particularly as large-scale projects like CMB-S4 begin to ramp up.

Efficient, effective-medium coatings for silicon lenses have been successfully demonstrated.
These coatings comprise sub-wavelength structures that are created by removing material from the lens surface \cite{datta:2013}.
While it is technically possible to use the same subtractive methods on an alumina lens, there are substantial practical hurdles to overcome.
Alumina's Vicker's hardness is almost twice that of single-crystal silicon (\SI{20.1}{\giga\pascal} vs. \SI{11.0}{\giga\pascal}) \cite{mccolm:1990}, which makes machining the material more difficult.
The equipment needed to cut or laser-ablate the surface of a \SI{\sim 70}{\centi\meter} diameter alumina lens currently only exists at commercial scale.
Outsourcing a large lens for laser ablation could be done, but the long processing time required makes that approach expensive.
One could build such a machine, but the technical expertise and time required to design, fabricate, verify, operate, and maintain it come at great cost.
Additionally, errors made during subtractive processes may be impossible to correct, which is a considerable risk when the base cost of the unmodified lens ($\sim$few$\,\times10^{4}$ USD and $\mathcal{O}(10)$ weeks lead time) is taken into account.

An alternative approach to the subtractive meta-material method is to layer the surface of the lens with materials that act as an AR coating.
We present two such coatings for alumina substrates in this paper: one for densely packed arrays of quasi-optical elements (henceforth ``lenslets''), the other for large-format lenses.
\autoref{sec:design} describes the coatings' design and fabrication; \autoref{sec:test} covers optical testing of the coatings.
We present the results of the optical tests in \autoref{sec:result} and discuss those results in \autoref{sec:discuss}.

\section{Coating design and fabrication}
\label{sec:design}
A common type of AR coating is a quarter-wavelength coating, which can be optimized for a single frequency, or narrow range of frequencies.
The material used for a coating of this type should have a refractive index equal to the geometric mean of the incident and substrate refractive indices, and its thickness should be one-quarter the in-material wavelength at the target frequency \cite{dobrowolski:2010}.
This principle can be extended to a coating composed of multiple dielectric layers, which broadens the effective bandwidth of the coating.

We used the multilayer dielectric approach because the design of the SPT-3G survey camera required the coating to provide coverage from \SIrange{\sim 77}{\sim 252}{\giga\hertz} \cite{spt3ginstrumentpaper}.
SPT-3G observes the CMB in three bandpasses centered at \SIlist{95;150;220}{\giga\hertz} with \SIlist{\sim 25; \sim 21; \sim 21}{\percent} fractional bandwidth at each respective frequency.
The bandpasses are designed to fall within atmospheric transmission windows where there is little atmospheric water vapor to attenuate the CMB radiation \cite{chamberlin:2012}.
We found the refractive indices of porous polytetrafluoroethylene (PTFE) in the form of Zitex G-115 ($n=1.234$), Porex PM-23J ($n=1.292$), Rogers Corporation RO3035 bondply ($n=1.679$), and Rogers Corporation RO3006 bondply ($n=2.249$) to be suitable for an alumina lens coating.
Each of the materials are PTFE matrices with non-PTFE inclusions that lower or raise the effective refractive index of the bulk material.
The Zitex and Porex have \si{\micro\meter}-scale pores that reduce the refractive index below that of raw PTFE; the Rogers materials include high-index ceramic particles, which boosts the overall refractive index.
The thickness of the RO3000-series bondply layers was dictated by their off-the-shelf availability: \SI{0.127}{\milli\meter} in both cases.
Though it is possible to specify an arbitrary thickness for the Zitex and Porex, we found the stock \SI{0.381}{\milli\meter} thick material to be suitable.

\begin{figure}[h!]
\centering
\subfloat[]{
  \includegraphics[width=0.475\textwidth]{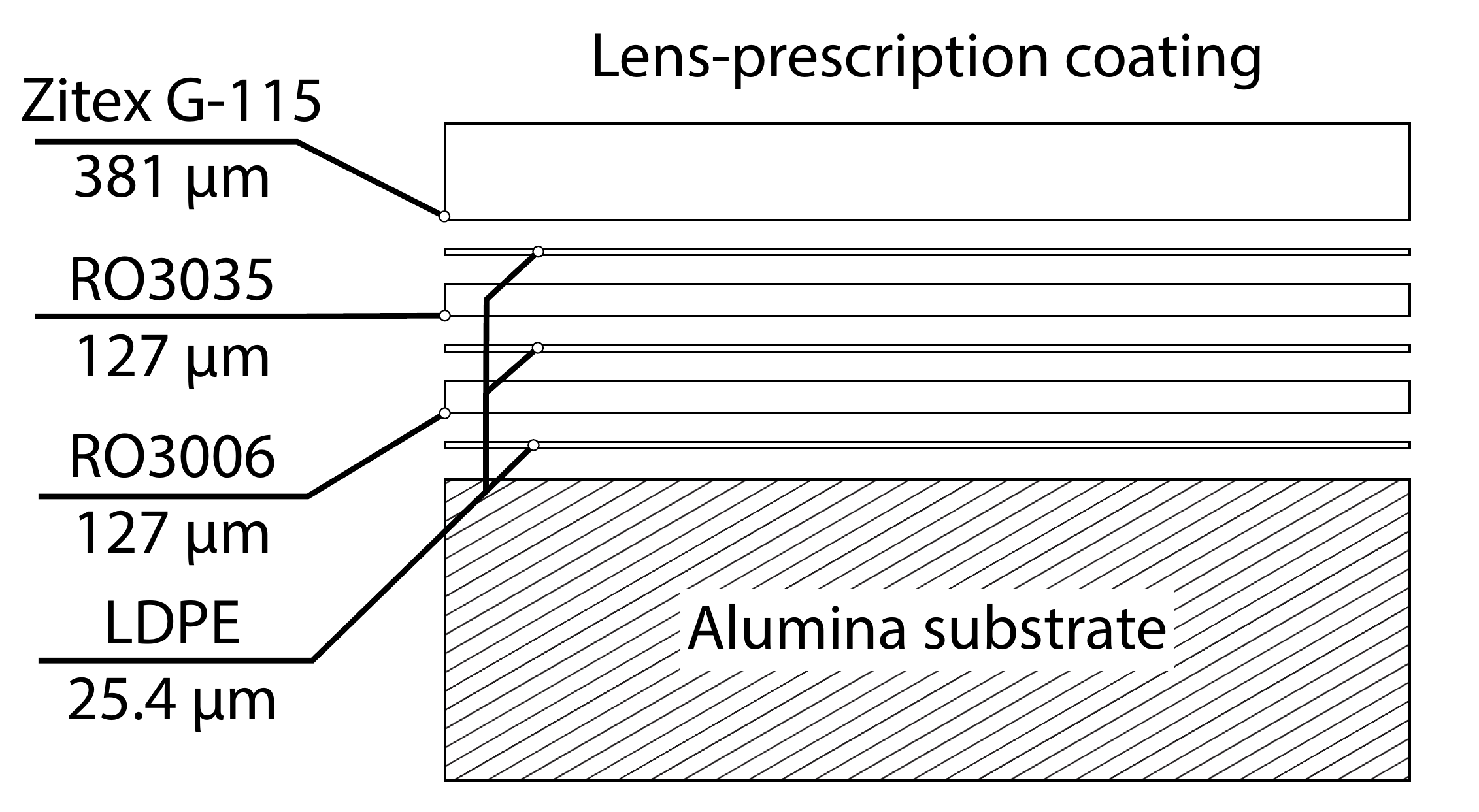}
}
\subfloat[]{
  \includegraphics[width=0.475\textwidth]{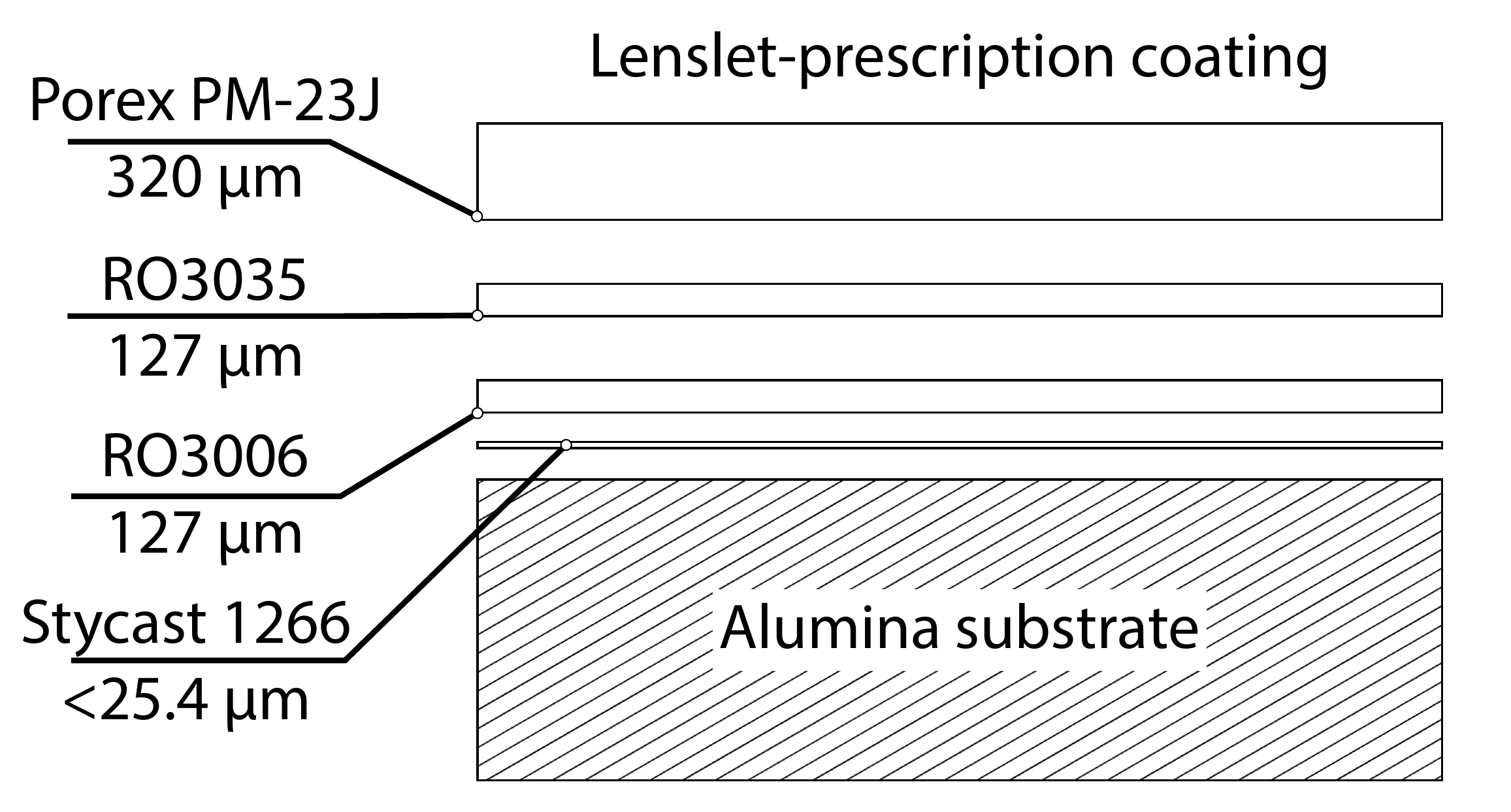}
}
\caption{
Diagrams of the SPT-3G lens- and lenslet prescription AR coatings showing the materials and nominal thickness of each layer:
\emph{(a)} The lens-prescription coating uses intermediate LDPE bonding layers between the main dielectric layers and the alumina substrate;
\emph{(b)} The main dielectric materials of the lenslet-prescription coating are self-bonded and require no intermediate LDPE layers.
It is bonded to the alumina substrate by a thin layer of Stycast 1266.
The manufactured thickness of the Porex layer is \SI{381}{\micro\meter}, but the layer is compressed during the bonding process.
The thickness shown is typical of that layer after bonding.
}
\label{fig:coatingdiagrams}
\end{figure}

\begin{figure}[h!]
\centering
\includegraphics[width=8.4cm]{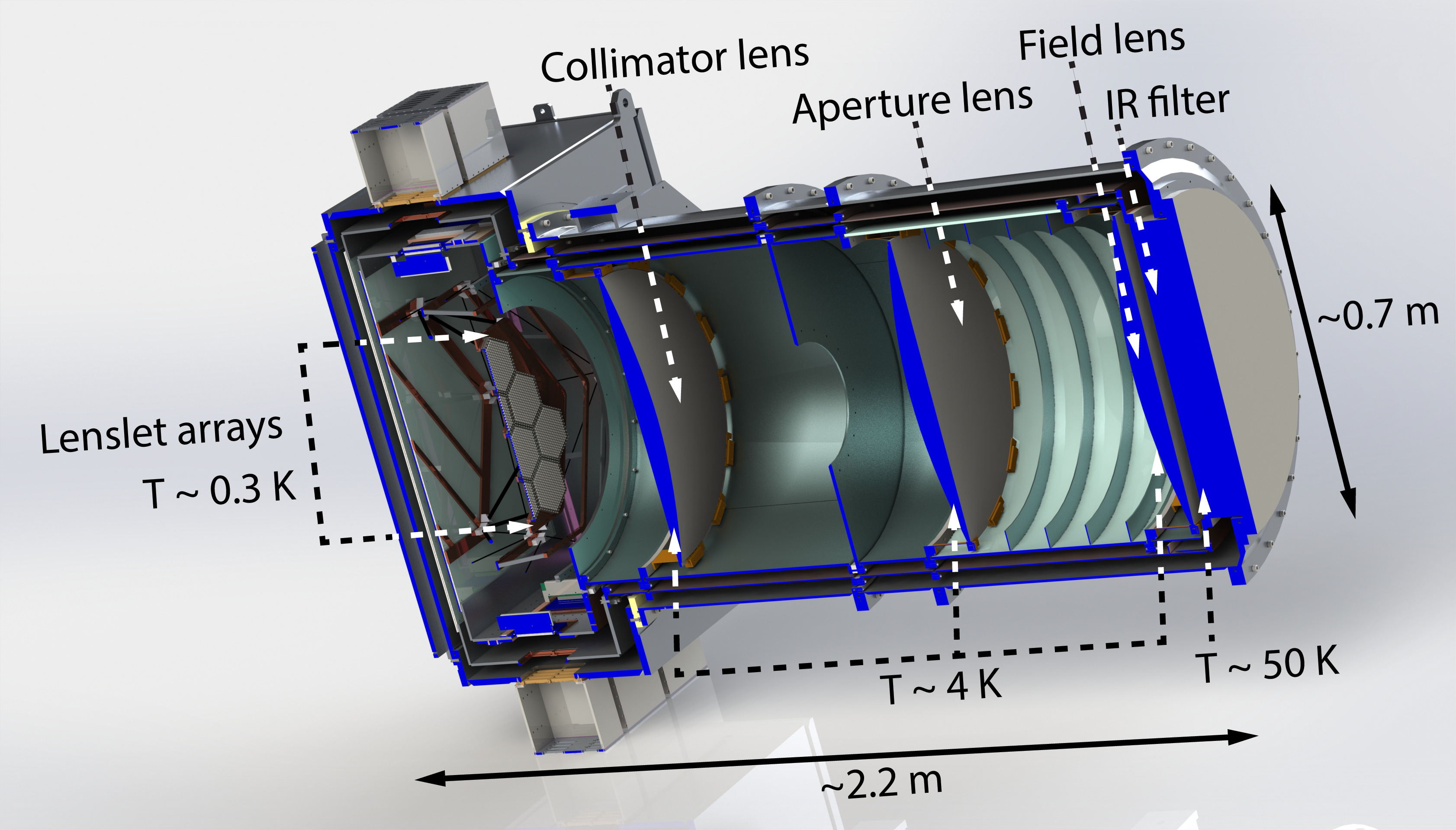}
\caption{Cutaway rendering of the SPT-3G cryostat showing the relative positions of the alumina infrared filter, refractive optics, and lenslets, as well as their approximate dimensions and operating temperatures.}
\label{fig:cryostatcutaway}
\end{figure}

We developed two coating prescriptions: one for large-format optics, and one for lenslet arrays.
\autoref{fig:coatingdiagrams} illustrates the layers and nominal thicknesses for both coating prescriptions.
The SPT-3G refractive optics and infrared filter are composed of CoorsTek AD-995 alumina; each element is \SI{\sim 720}{\milli\meter} in diameter.
The infrared filter operates at \SI{\sim 50}{\kelvin}, while the refractive optics operate at \SI{\sim 4}{\kelvin}.
The lenslets are hemispheres of \SI{99.5}{\percent} pure alumina with \SI{5}{\milli\meter} outer diameter.
They operate at \SI{\sim 300}{\milli\kelvin}, coupling CMB radiation that has passed through the refractive optics to the instrument's detectors (\autoref{fig:cryostatcutaway}).
The raw lenslets were produced by Kyocera and then ground to the hemispherical specification by TN Michigan (Hoover Precision at the time of manufacture).

\begin{figure}[h!]
\begin{center}
\includegraphics[width=8.4cm]{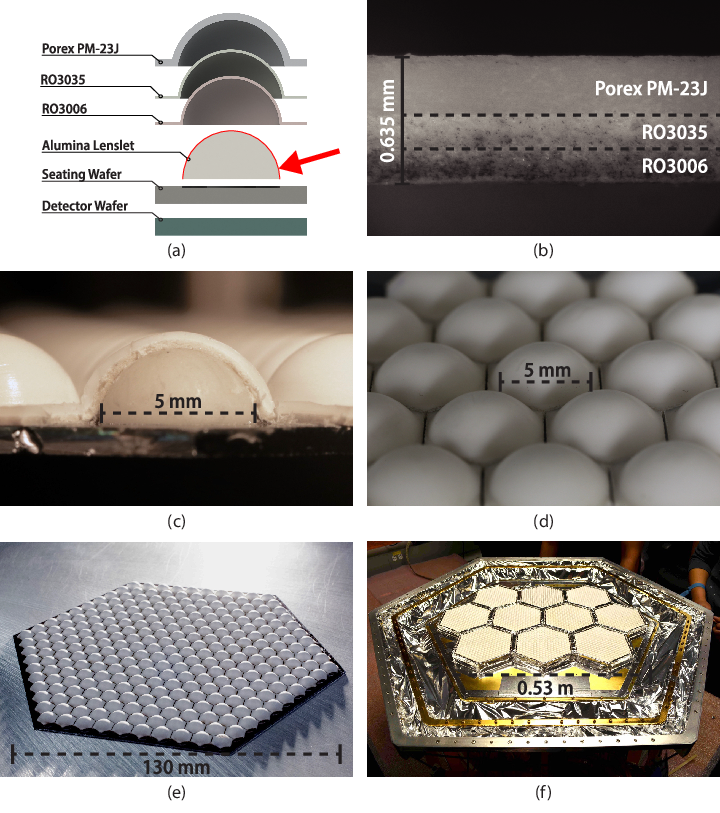}
\end{center}
\caption{
\emph{(a)} Exploded-view schematic of the lenslet-prescription AR coating, lenslet, lenslet seating wafer, and detector wafer.
The thin, red line called out by the red arrow marks the position of the Stycast 1266 layer;
\emph{(b)} Infrared photograph of a lenslet AR coating after lamination.
Dashed lines mark the approximate boundaries of the original materials;
\emph{(c)} Cutaway of a lenslet AR coating after the molding and Stycast procedures.
The coating and lenslet surface were marred during cross-sectioning;
\emph{(d)} Close-up photograph of an assembled, AR-coated, and laser-diced lenslet array;
\emph{(e)} Photograph of an AR-coated lenslet array. The array comprises 271 coated elements;
\emph{(f)} Photograph of the SPT-3G detector array taken before its final integration with the detector cryostat.
The ten white hexagons are AR-coated lenslet arrays.
}
\label{fig:lensletcycle}
\end{figure}

The lenslet-prescription coating is self-bonded, requiring no adhesive layer between the dielectric layers (\autoref{fig:coatingdiagrams}b and \autoref{fig:lensletcycle}a).
The layers are clamped together, then baked in a PID-controlled oven.
At \SI{\sim 371}{\degreeCelsius} the PTFE molecules begin to coalesce and diffuse across layer boundaries.
When the oven is cooled, the interlinked PTFE molecules crystallize, yielding a single sheet with three dielectric subsections (\autoref{fig:lensletcycle}b).
Clamping and baking the coating typically decreases the thickness of the coating by \SI{\sim 60}{\micro\meter} (\SI{\sim 15}{\percent}).
The RO3035 and RO3006 layers are virtually incompressible, so all of the change in thickness occurs in the low-index Porex layer---this is due to pore collapse within the material.
The decrease in thickness is offset by an increase in refractive index.
In the limit that the Porex becomes fully dense, it should assume the refractive index of pure PTFE ($n=1.44$).
Once the laminated coating cools to room temperature, it is molded to fit the lenslet array by a die press.
A calibrated volume of Stycast 1266 epoxy is then deposited at the apex of each lenslet, and the coating is secured to the lenslet array.
After the epoxy cures, the coating is laser-diced\endnote{Epilog Zing 16, \SI{30}{\watt}: \url{https://www.epiloglaser.com/laser-machines/zing-engraver-cutter/}} to reduce thermal stress during the cooldown to \SI{300}{\milli\kelvin}.
The laser cuts a hexagonal pattern around each lenslet, separating it from its neighbors (\autoref{fig:lensletcycle}c--e).

\begin{figure}[h!]
\centering
\includegraphics[width=8.4cm]{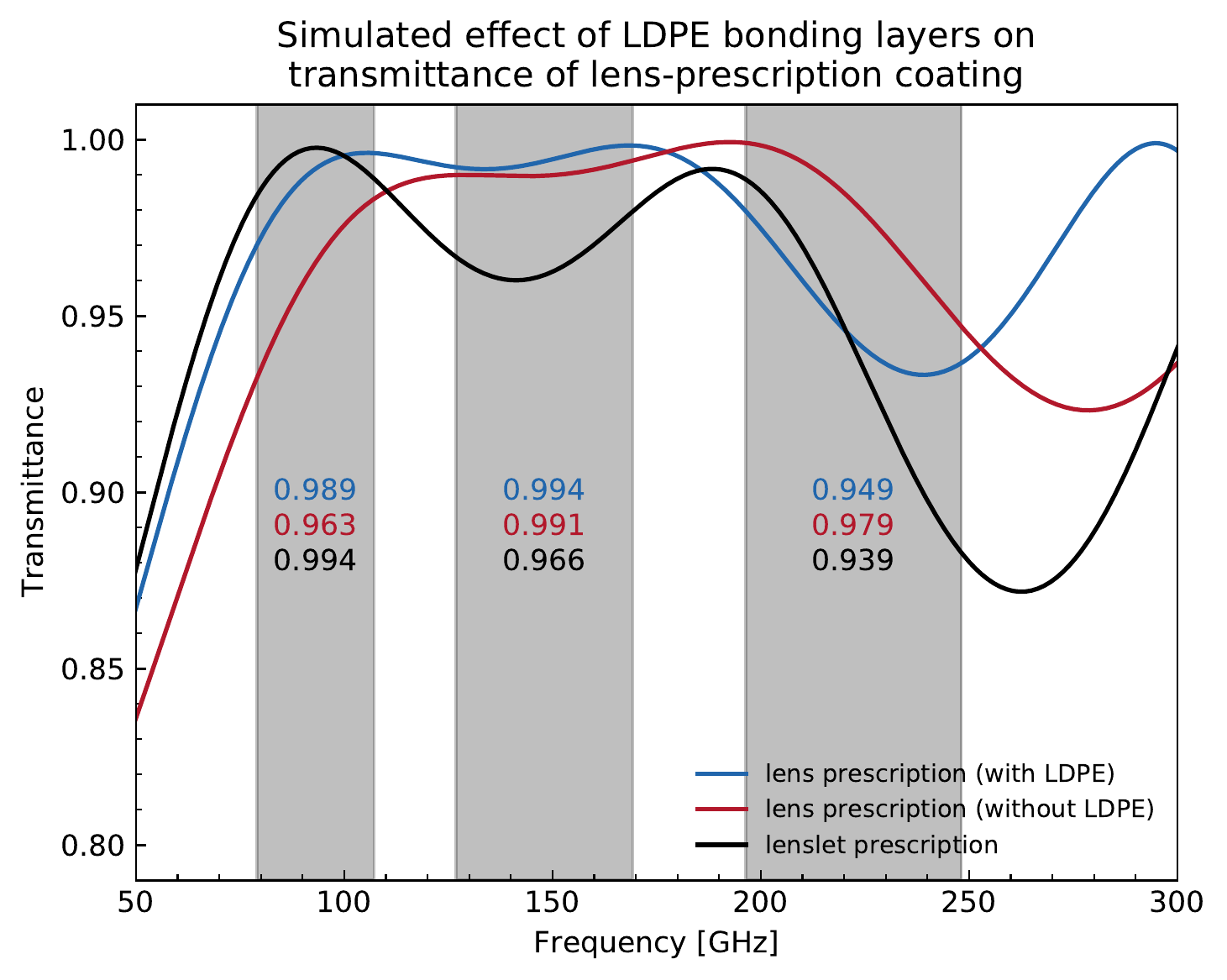}
\caption{
Modeled transmittance of a single-surface, lossless lens-prescription coating with \emph{(blue)} and without \emph{(red)} LDPE bonding layers on a semi-infinite lossless alumina substrate.
The modeled response of a lenslet-prescription coating \emph{(black}), which is self-bonded and does not use LDPE, is shown for comparison.
The refractive indices used in the simulations are the \SI{300}{\kelvin} measured values from \autoref{table:matprop}, except for that of LDPE which is adopted from \cite{lamb:1996}.
The difference between the lens-prescription coating without LDPE and the lenslet-prescription coating lies in the low-dielectric layer: the lens prescription uses Zitex G-115; the lenslet prescription uses Porex PM-23J.
The vertical grey bars mark the SPT-3G \SIlist{95;150;220}{\giga\hertz} observing bands, respectively.
LDPE bonding layers affect the response of the coating at the level of a few percent, boosting transmittance in the \SIlist{95;150}{\giga\hertz} bands and depressing it in the \SI{220}{\giga\hertz} band.
The mean in-band transmittance is given for each band in blue \emph{(top)}, red \emph{(middle)}, and black \emph{(bottom)} text, corresponding to simulations of the lens-prescription coating with LDPE, the lens-prescription coating without LDPE, and the lenslet-prescription coating, respectively.}
\label{fig:compareldpe}
\end{figure}

The lens-prescription coating (\autoref{fig:coatingdiagrams}a) uses Zitex G-115 instead of Porex PM-23J as the low-index layer.
For our purposes, the effective differences between the Zitex and Porex are small enough that we opt to use the lower cost Zitex for the large-format optics.
The lens-prescription coating also incorporates \SI{\sim 25.4}{\micro\meter} thick low-density polyethylene (LDPE) as an adhesive layer between the dielectric layers, instead of self-bonding and Stycast 1266 like the lenslet prescription.
This is largely due to practical difficulties in self-bonding large coating sections and accurately aligning coating sections during layout.
Simulations show the LDPE bonding layers slightly increase transmittance in the \SI{95}{\giga\hertz} and \SI{150}{\giga\hertz} bands while slightly reducing transmittance in the \SI{220}{\giga\hertz} band (\autoref{fig:compareldpe}).
The coating is assembled one layer at a time by a vacuum-bagging process.
The work piece is sealed in a mold from which the air is then evacuated.
Atmospheric pressure outside the mold provides the force needed to laminate the coating layers.
Each layer is baked at \SI{140}{\degreeCelsius} during its initial layout.
We find that once all layers are applied, a final bake at a soak temperature of \SI{160}{\degreeCelsius} results in robust lamination.

\begin{figure}[h!]
\centering
\includegraphics[width=8.4cm]{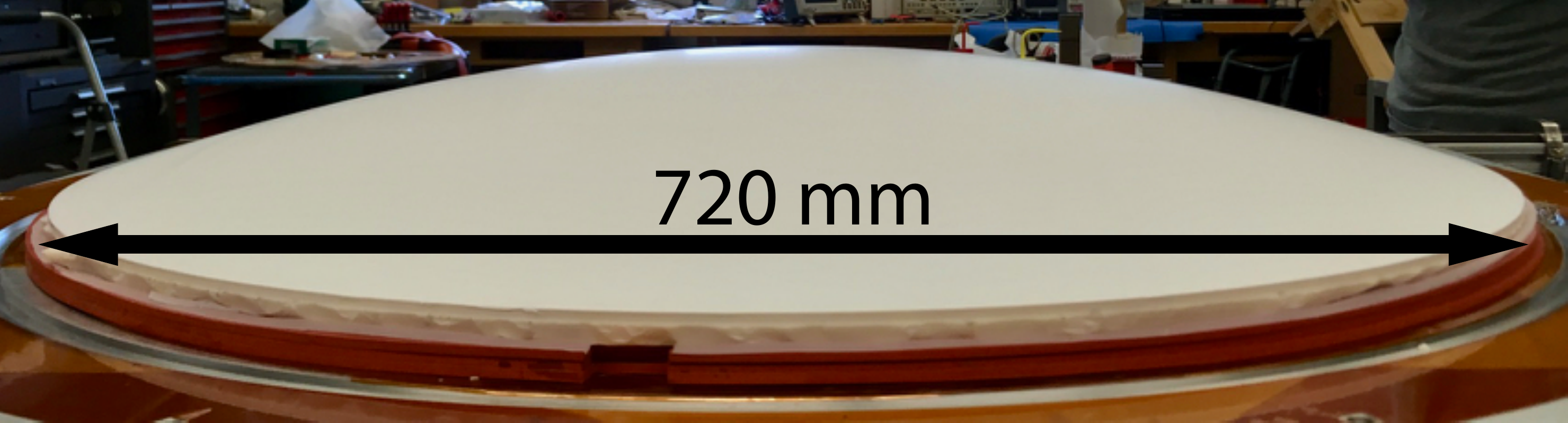}
\caption{Photograph of the AR-coated SPT-3G collimator lens illustrating its radius of curvature.
The lens is \SI{\sim 69}{\milli\meter} thick at its center.}
\label{fig:collimatorprofile}
\end{figure}

\begin{figure}[h!]
\centering
\subfloat[]{
  \includegraphics[width=0.22\textwidth]{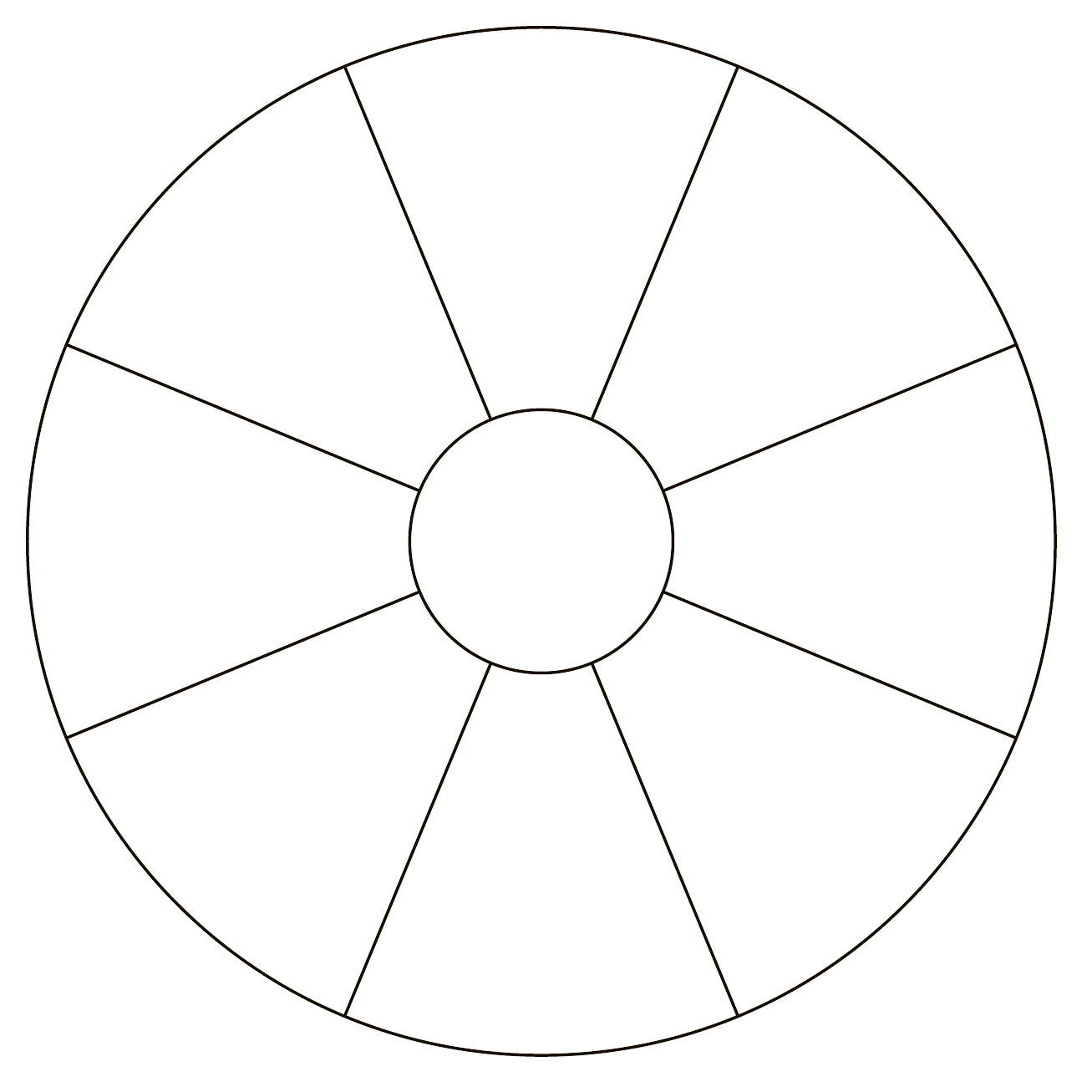}
}
\subfloat[]{
  \includegraphics[width=0.22\textwidth]{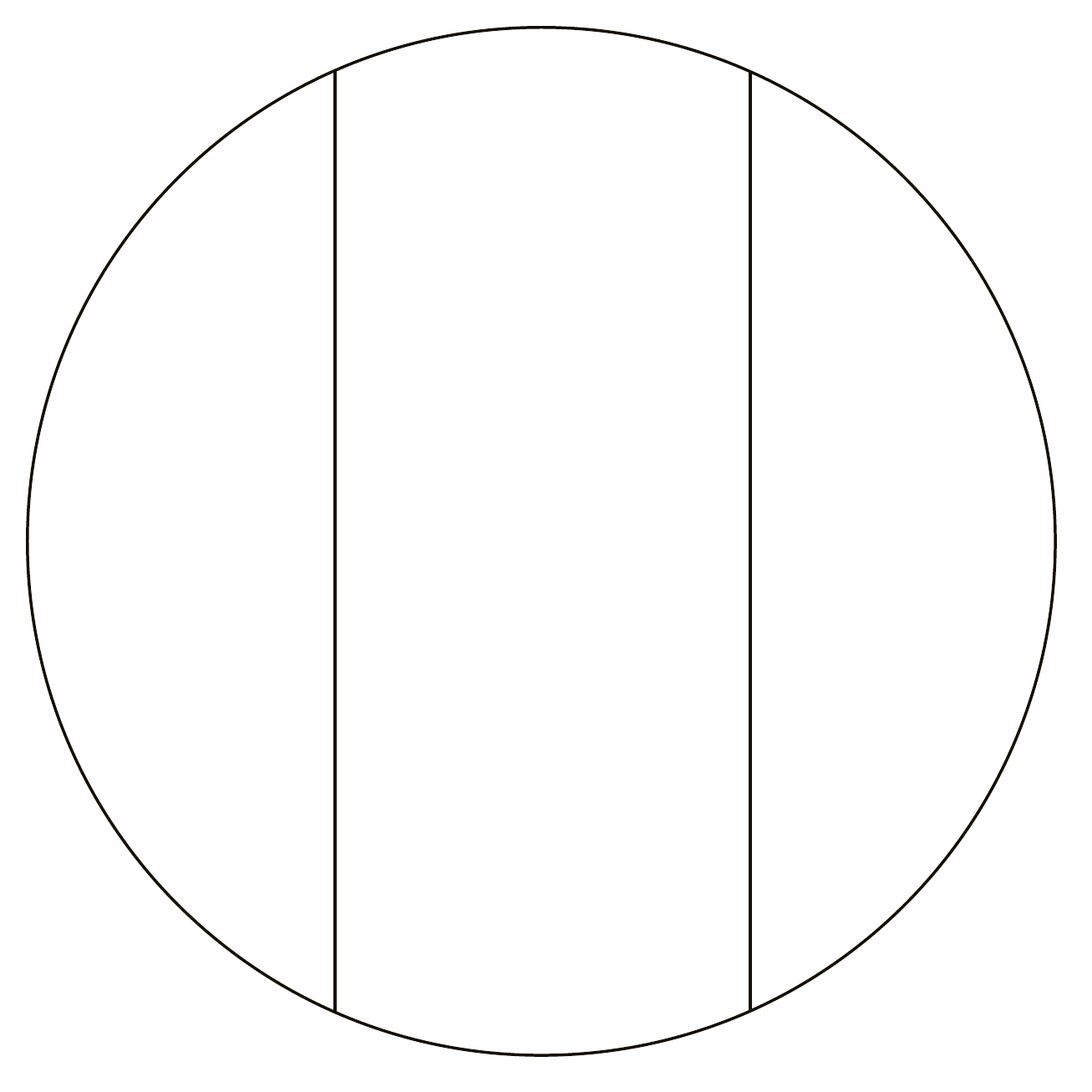}
}
\subfloat[]{
  \includegraphics[width=0.22\textwidth]{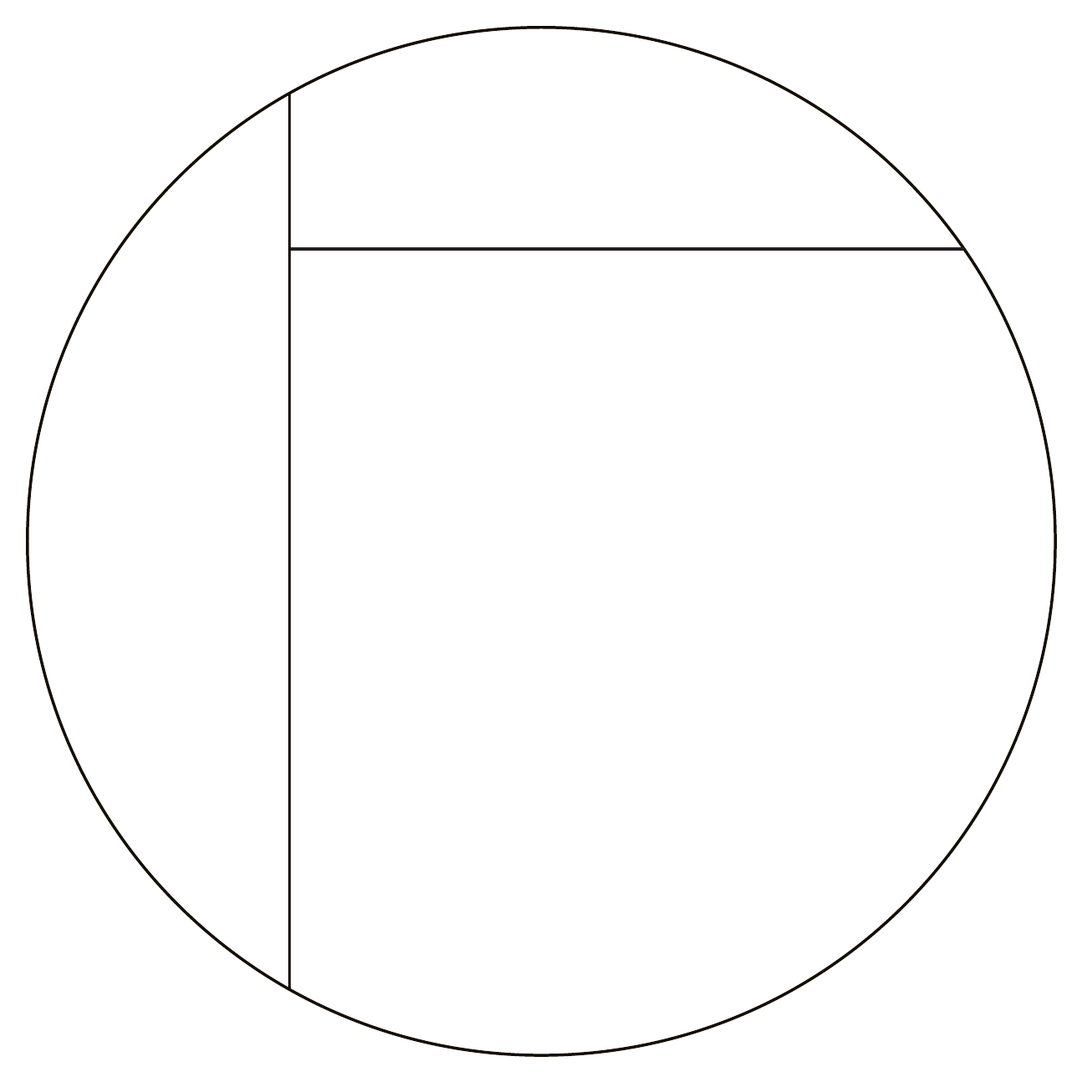}
}
\caption{
We chose the illustrated seam patterns in an effort to minimize AR coating material waste while covering the largest possible area without the material wrinkling.
Zitex G-115 is very compliant, so we were able to use it in its manufactured form.
The RO3035 and RO3006 materials, however, are somewhat stiffer and prone to tearing.
For this reason we selected a radially symmetric petal-like pattern for the curved surfaces.
The RO3035 and RO3006 seam pattern for flat surfaces provided the most efficient practical use of the materials.
\emph{(a)} Seam layout for the RO3035 and RO3006 layers on curved surfaces.
Sections were clocked and scaled as necessary to minimize seam overlap;
\emph{(b)} Seam layout for the Zitex G-115 layer on both flat and curved surfaces;
\emph{(c)} Seam layout for the RO3006 and RO3035 layers on flat surfaces.
Sections were clocked to minimize seam overlap.
Projections of the seams for both the lenses and infrared filter are shown in \autoref{fig:seamproj}.
}
\label{fig:lenspattern}
\end{figure}

The scale of the lenses causes additional complexities: the mismatch between alumina and PTFE's coefficients of thermal expansion can result in coating delamination; there is also the problem of wrapping a curved surface with a flat sheet (\autoref{fig:collimatorprofile}) \cite{demaine:2009}.
It is not possible to coat the entire optical surface of the large-format SPT-3G optics without seams, in part due to the form factor of the raw materials.
Therefore we opt for seam patterns that allow us to use the AR coating materials most efficiently and avoid unnecessary waste (\autoref{fig:lenspattern}).
\begin{figure}[h!]
\centering
\subfloat[]{
  \includegraphics[width=0.3\textwidth]{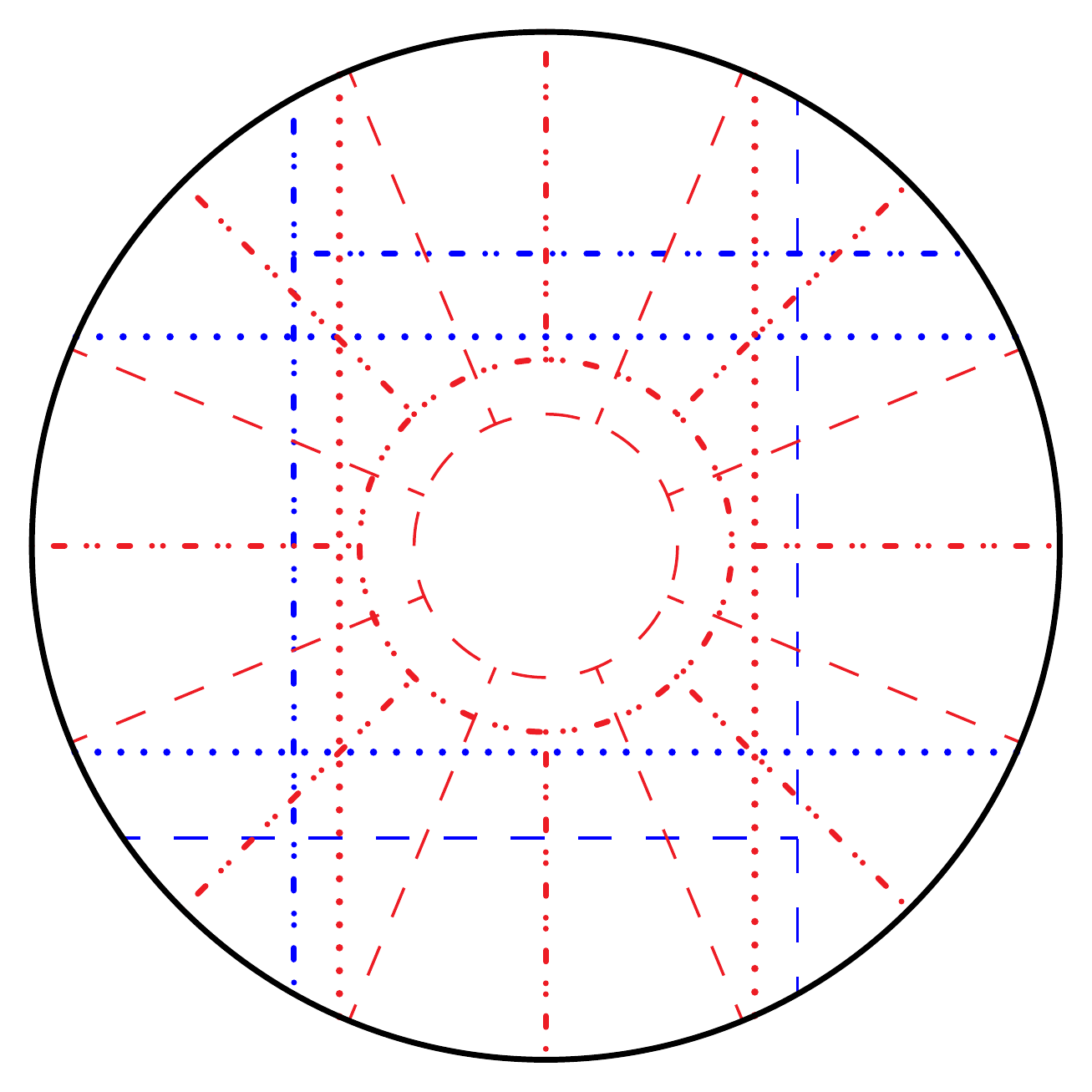}
}
\subfloat[]{
  \includegraphics[width=0.3\textwidth]{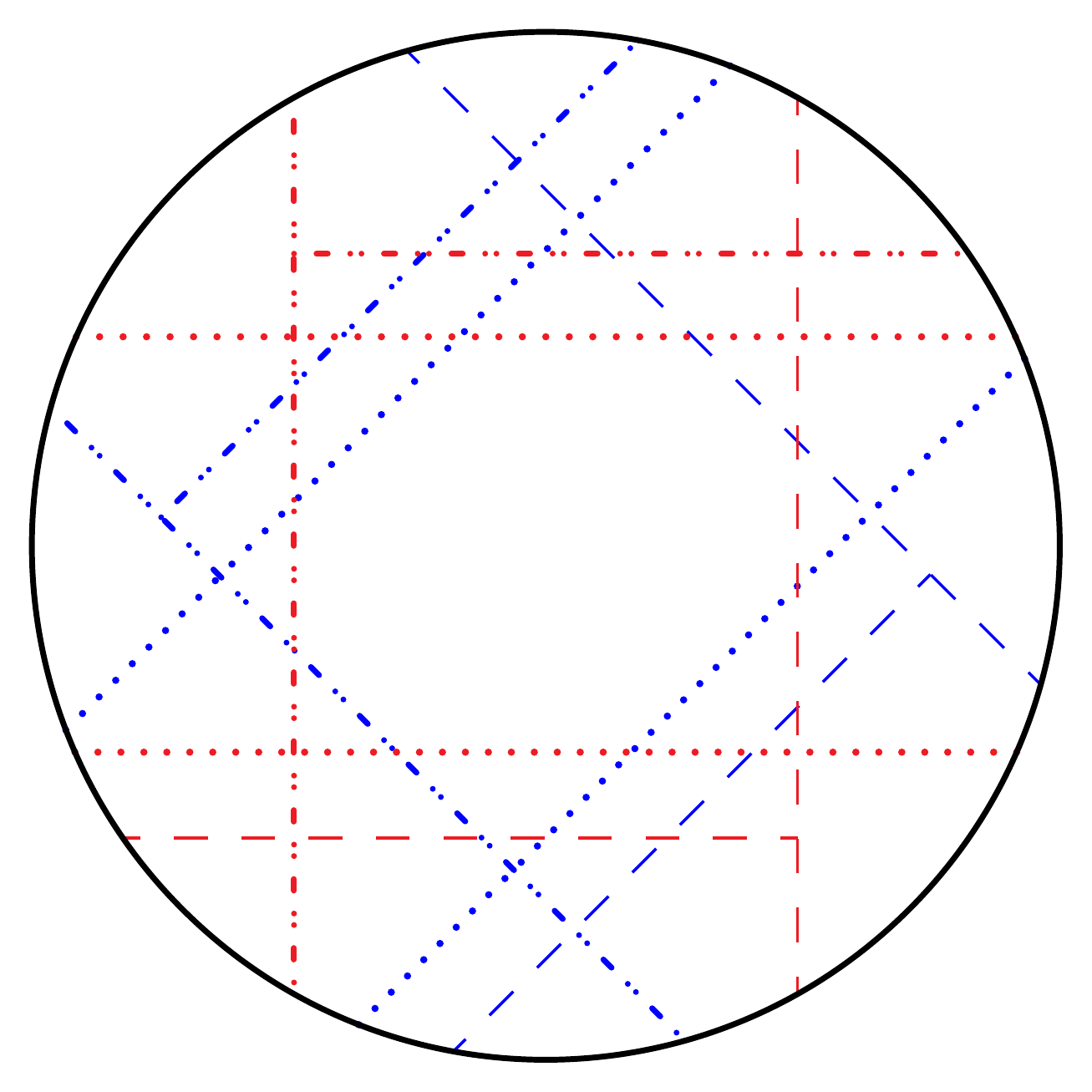}
}
\caption{
Projections of the seam layout for the three coated SPT-3G lenses \emph{(a)} and the single coated infrared filter \emph{(b)} onto a plane parallel with the optics' flat surfaces.
The projections show all three coating materials on both sides of the large-format optics: Zitex G-115 \emph{(dotted)}, RO3035 \emph{(dashed)}, and RO3006 \emph{(dot-dot-dashed)}.
Flat/bottom surfaces are marked in \emph{blue} and curved/top surfaces are marked in \emph{red}.
}
\label{fig:seamproj}
\end{figure}
Zitex-G115 is a pliable material and lays across flat and curved surfaces without issue, so we use it without modification (other than trimming it to width where necessary).
The RO3035 and RO3006 materials are somewhat stiffer and are prone to tearing if they wrinkle, which is a problem when attempting to lay out a full sheet over a curved lens surface.
We choose to use a radially symmetric petal-like pattern on curved surfaces for the RO3035 and RO3006 layers to avoid wrinkling or tearing.
On flat surfaces we use a pattern that minimizes waste of the Rogers Corporation materials.
Projections of the seam layout onto a plane parallel to the optics' flat surfaces are shown in \autoref{fig:seamproj}.
Additional details regarding the manufacture of both coatings, including production time tables, are given in \cite{nadolski:2018}.

The seams in each AR coating layer are \SI{< 100}{\micro\meter} wide and conservatively estimated to occupy \SI{< 0.5}{\percent} of the total optical area of a lens.
Even at the high-frequency edge of the \SI{220}{\giga\hertz} SPT-3G observing band, $D_{seam} / \lambda < 0.1$.
Therefore, we anticipate scattering loss due to the seams to be dominated by reflective and absorptive loss due to the AR coating and bulk alumina, respectively.
We expect the AR coating seams will cause sidelobes in the SPT beam because they will appear as \si{\milli\meter} scale discontinuities at the primary reflector.
The primary reflector itself is composed of \num{218} panels separated by \SI{\sim 1}{\milli\meter} gaps, which are filled by low-profile beryllium-copper spring flanges \cite{padin:2008,padin:2008b,carlstrom:2011}.
These discontinuities are known to cause sidelobes due to scattering, and their effects on the PSF have been simulated.
We plan to include the effects of the AR coating seams in a new study of primary reflector sidelobes, but that analysis is outside the scope of this work.

\section{Experimental procedure}
\label{sec:test}
We used a Winston cone-coupled, cryogenic bolometer and Michelson Fourier transform spectrometer (FTS) to characterize the performance of the lens- and lenslet-prescription coatings and their constituent materials between \SIlist{150;350}{\giga\hertz} at \SIlist{\sim 300;\sim 77}{\kelvin}.
The monolithic silicon bolometer is described in \cite{downey:1984}; the FTS design is described in \cite{shoemaker:1980}, and is similar to the COBE/FIRAS design \cite{mather:1993}.
We conducted two rounds of sample-in/sample-out measurements to determine the transmittance of the samples.
First we measured the uncoated sample substrates to establish a baseline response for each substrate.
Then we coated the substrates and measured the samples twice more: once warm, and once cold.
A more detailed explanation of the sample preparation and measurement follows in the next subsections.

\subsection{Sample preparation}
\label{subsec:prep}
Taken by themselves, the individual AR coating materials were too thin to yield useful transmission spectra.
Therefore, we used CoorsTek AD-995 alumina discs (\SI{152.4}{\milli\meter} outer diameter $\times$ \SI{6.35}{\milli\meter} thick) as a mounting substrate to produce Fabry-Perot-like interference spectra.
The discs were all members of the same powder and firing batch, so we did not expect significant variations in their properties.
Nevertheless, we measured the transmission spectra of each bare alumina disc in order to establish a baseline and decouple the substrate properties from those of the coating materials.

We assigned each substrate a coating material to be laminated to the substrate by a vacuum-bagging process, using \SI{\sim 25.4}{\micro\meter} thick LDPE as the bonding agent.
Before laminating the samples, we measured the thickness of every layer of every sample at 20 positions spread across its \SI{\sim 150}{\milli\meter} diameter surface using a digital micrometer perpendicularly mounted to a granite flat-surface plate.
Doing so allowed us to check for thickness gradients across the layers (we saw no evidence of this) and provided an average layer thickness that we used later during the data analysis phase.

Both sides of each substrate were coated by a single, stock-thickness layer of dielectric material, with the exception of two discs that were prepared using the complete, multilayer lens- and lenslet-prescription coatings.
In total, we prepared six sample discs: Zitex G-115, Porex PM-23J, Rogers RO3035 bondply, Rogers RO3006 bondply, lenslet prescription, and lens prescription.
We also had one \SI{25.4}{\milli\meter} thick alumina sample that we did not coat, the purpose of which was to characterize attenuation in the bulk alumina.
The thick alumina sample came from a different powder and firing batch than the coated samples.

\subsection{Experiment set up}
\label{subsec:exp}
\begin{figure}[h!]
\centering
\includegraphics[width=8.4cm]{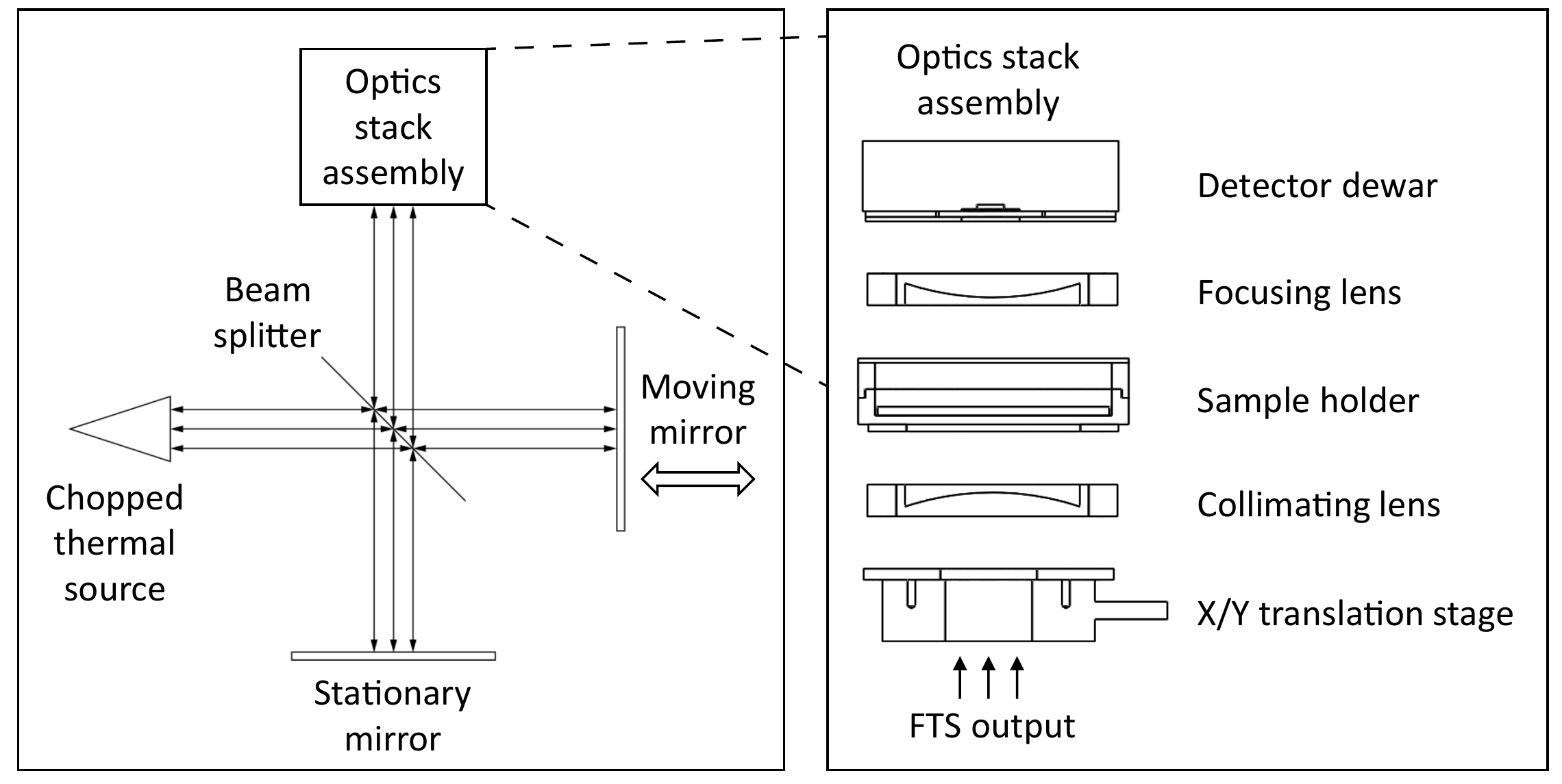}
\caption{Conceptual diagram of the experiment's optical chain \emph{(left)} and a detailed diagram of the optics stack assembly \emph{(right)}.
Eccosorb HR-10 (not shown) was attached to surfaces of the optics stack assembly along the optical path in order to minimize stray reflections.
The liquid helium-cooled detector dewar contains a monolithic silicon bolometer that is coupled via a Winston cone.
Dimensions are not to scale.}
\label{fig:ftsdiagram}
\end{figure}

We used a chopped thermal source as the input signal to the FTS.
The output of the FTS was collimated through the test sample and then focused onto a cryogenic bolometer by two nylon lenses ($n=1.727$; \autoref{fig:ftsdiagram}) \cite{lamb:1996}.
The bolometer was read out by an analog-to-digital converter and data acquisition software.
We minimized stray reflections by covering surfaces along the optical path with Eccosorb HR-10, a strong \si{\milli\meter} wave absorber.

Our procedure for a measurement was to conduct a reference scan set with a clear optical path (sample-out), then a test scan set with the sample placed in the optical path (sample-in).
Each scan set produced between three and nine interferograms.
We calculated the Fourier transform of each interferogram in a scan set, then averaged those spectra to obtain the averaged spectrum for the set.
We then calculated the transmittance spectrum by taking the ratio of the sample-in average spectrum to the sample-out average spectrum.

\subsection{Additional measurements}
\label{subsec:reflectanceexperiment}
The reflectance of a lens-prescription sample was measured using a reflectometer at the University of Michigan; the reflectometer design is described in \cite{chesmore:2018}.
The sample was measured at a \SI{10}{\degree} angle of incidence between \SIlist{\sim 90;\sim 180}{\giga\hertz}.
Though the AR coating was prepared in the same manner as above, the reflectometer-tested sample employed an alumina substrate of different formulation to CoorsTek AD-995.

We also performed a vector network analyzer (VNA; Keysight 8510C) single-port S11 measurement of the same lens-prescription sample used in the FTS tests.
The VNA signal was coupled to free space by a WR-10 rectangular horn antenna, and collimated by a \SI{90}{\degree} refocusing mirror.
n the same manner as \cite{barkats:2018}.
The sample was placed at a position where it filled the beam.
The normal-incidence S11 measurement ranged from \SIrange{75}{110}{\giga\hertz} and was calibrated to an aluminum mirror (short) and mm-wave absorber (load).
We employed time-domain gating to minimize spurious signals from within the room and at interfaces throughout the waveguide set-up leading to the feedhorn.

\section{Results}
\label{sec:result}
The cryogenic bolometer used in the FTS tests was responsive above \SI{\sim 130}{\giga\hertz} and low-pass filtered at \SI{\sim 600}{\giga\hertz}, so we restricted our analysis to frequencies between \SIlist{150;350}{\giga\hertz}, where the uncertainties are smallest.
We expect the measurement uncertainties to be dominated by statistics rather than systematics: the change in detector response over the course of a full sequence of reference and test measurements (\SI{< 10}{\minute}) was negligible, and other systematics were eliminated due to the sample-in/sample-out measurement strategy.

In our analysis, we first extracted the refractive index of the uncoated, \SI{6.35}{\milli\meter} thick alumina samples by modelling them as lossy dielectric slabs following \cite{halpern:1986}.
Due to the low loss tangent of alumina, measurements of these thin substrates were not a good probe of the material's loss.
Therefore, we used measurements of the \SI{25.4}{\milli\meter} thick alumina sample to constrain the loss tangent of the material.
Placing limits on the alumina optical properties allowed us to better decouple the effects of the alumina from those of the AR coating plastics.

\begin{figure}[h!]
\centering
\includegraphics[width=8.4cm]{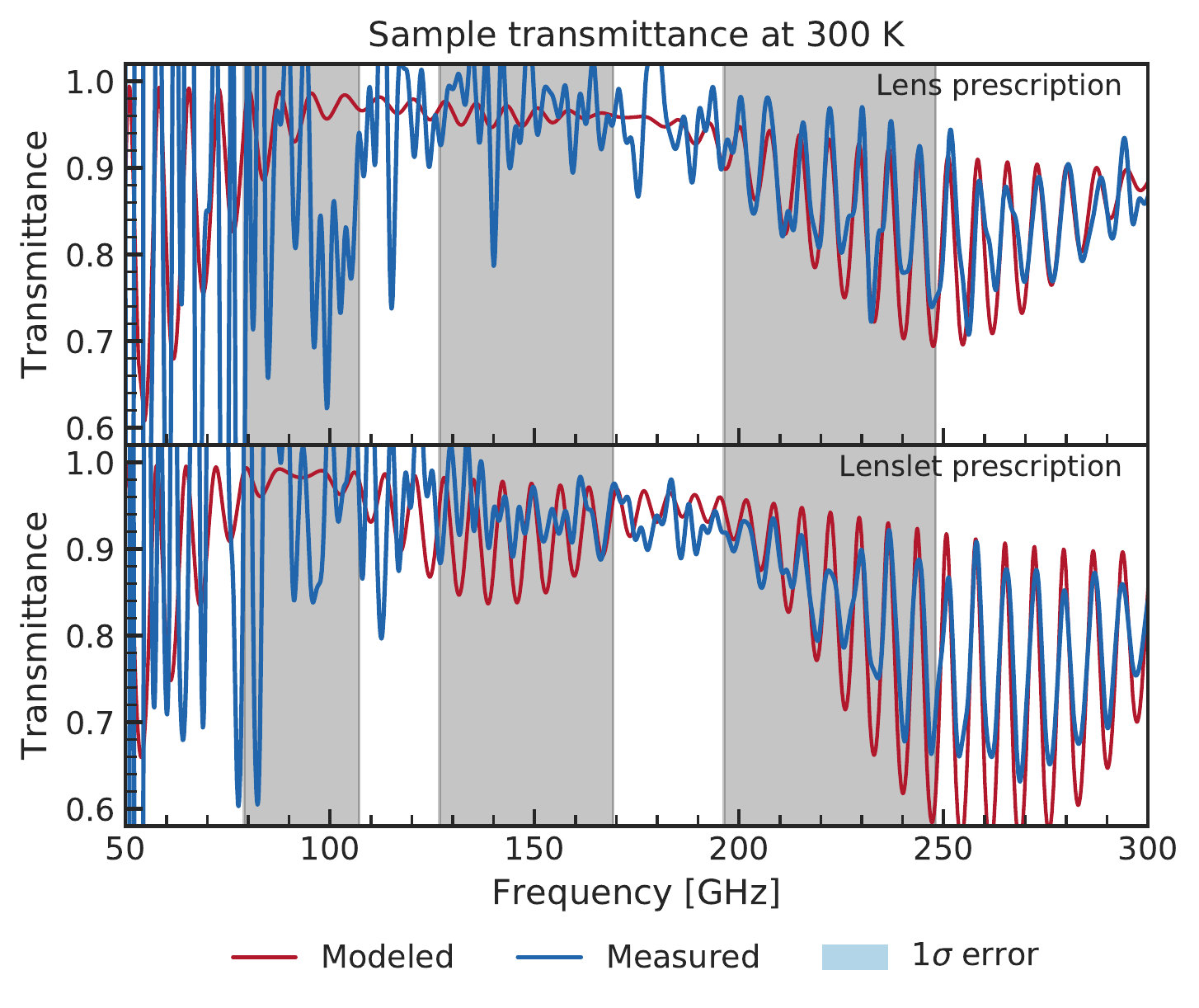}
\caption{
Measured transmittance of the lens- \emph{(top)} and lenslet-prescription coatings \emph{(bottom)} at room temperature.
The solid blue line and light-blue shaded region are the measurement data and associated $1\sigma$ uncertainties; the red line is the model, which is produced using best-fit values for the optical properties of each material.
The vertical grey bars mark the SPT-3G \SIlist{95;150;220}{\giga\hertz} observing bands, respectively.
The passband of the detector used in the experiment was \SIrange{\sim 130}{\sim 600}{\giga\hertz}, but we restricted the fit region to \SIrange{150}{350}{\giga\hertz}. 
The lens-prescription data curve and associated error bars are the result of five interferograms, while the lenslet-prescription data curve and associated error bars are the result of eight.}
\label{fig:coatingtransmittance}
\end{figure}

\begin{figure}[h!]
\centering
\includegraphics[width=8.4cm]{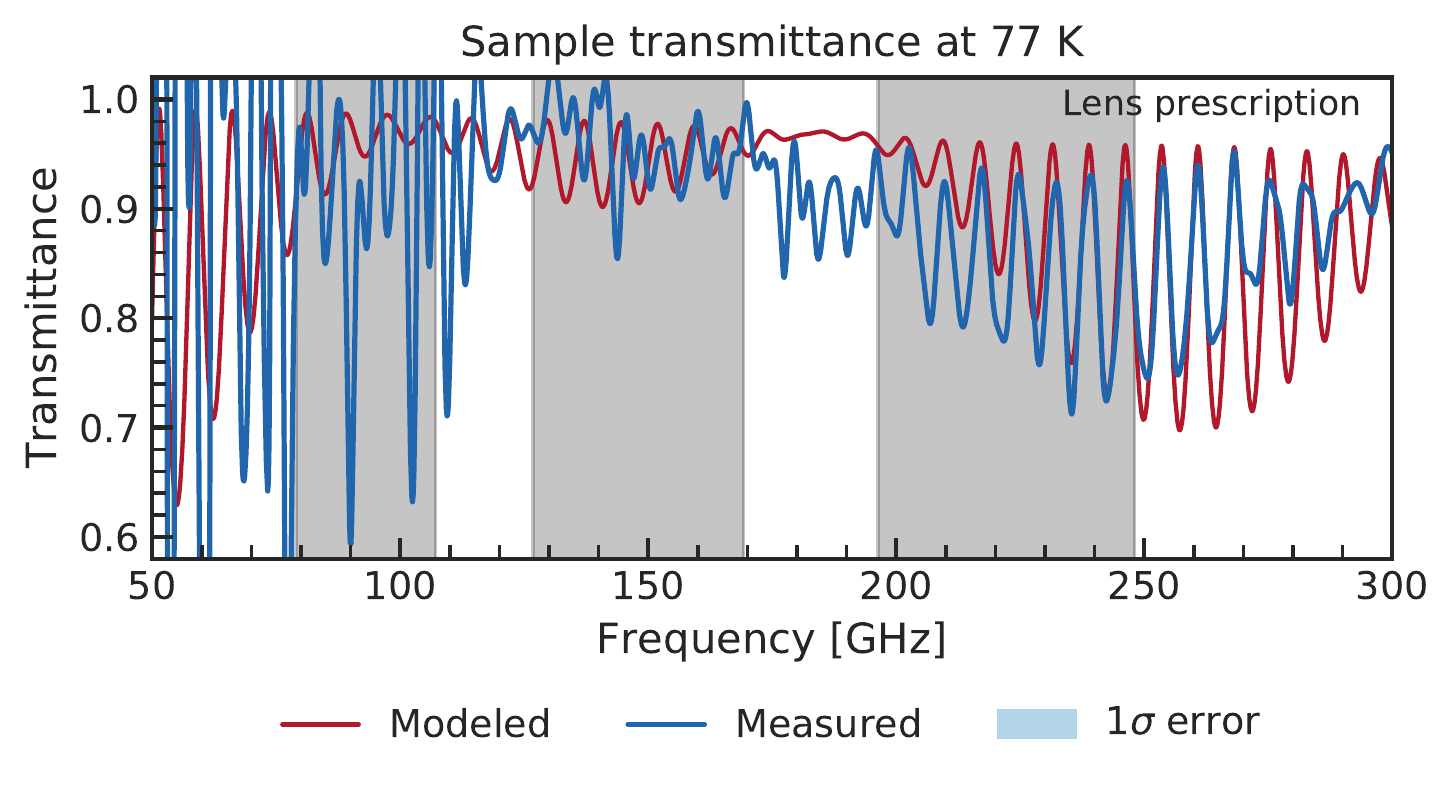}
\caption{
Measured transmittance of the lens-prescription coating at \SI{77}{\kelvin}.
We conductively cooled the sample through the edges of the discs using liquid nitrogen.
This allowed the interior of the sample holder (above the surface of the sample and liquid-phase nitrogen) to be a predominantly nitrogen gas environment, preventing frost build-up on the sample.
The solid blue line and light-blue shaded region are the measurement data and associated $1\sigma$ uncertainties; the red line is the model, which is produced using best-fit values for the optical properties of each material.
The vertical grey bars mark the SPT-3G \SIlist{95;150;220}{\giga\hertz} observing bands, respectively.
The passband of the bolometer used in the experiment was \SIrange{\sim 130}{\sim 600}{\giga\hertz}, but we restricted the fit region to \SIrange{150}{350}{\giga\hertz}. 
The lens-prescription data curve and associated error bars are the result of nine interferograms.
As discussed in \autoref{sec:discuss}, the lenslet-prescription coating used for testing was a much larger diameter than the the coatings currently deployed in the SPT-3G survey camera (\SI{\sim 150}{\milli\meter} vs. \SI{\sim 5}{\milli\meter}).
The tested lenslet-prescription coating did not survive the cooldown to \SI{77}{\kelvin} because we did not take our usual steps to mitigate the CTE mismatch between the coating and substrate, hence we do not show data for that sample.
}
\label{fig:coatingtransmittancecold}
\end{figure}

We used the characteristic matrix method to fit for the refractive index and loss tangent of the AR coating materials in the single-layer samples, using the previously determined alumina refractive index and loss tangent, the measured layer thicknesses, and an assumed refractive index and loss tangent for the LDPE bonding layer as model constraints \cite{heavens:1955,hou:1974}.
We adopted the LDPE refractive index and loss tangent from \cite{lamb:1996} that best matched the measurement's frequency and temperature conditions.
We used the resulting fit values for the AR material porperties (i.e., the inferred refractive indices and loss tangents) and adoptive LDPE values to simulate the response of the lens- and lenslet-prescription coatings, which we then compared to measurements (\autoref{fig:coatingtransmittance} and \autoref{fig:coatingtransmittancecold}).
We accounted for changes in material thickness due to thermal contraction at \SI{77}{\kelvin} using CTE values from existing literature \cite{wachtman:1962,lamb:1996,tonkin:1996,blumm:2010}.
At \SI{77}{\kelvin} the PTFE layers shrink by \SI{< 2}{\percent}, and the effect on $n$ is small compared to uncertainties in the cold measurement; changes in the alumina at \SI{77}{\kelvin} are even smaller.

We report the inferred material properties of the AR coating materials in \autoref{table:matprop}.
In \autoref{table:coateff1} we report the average transmittance and associated $1\sigma$ uncertainties of the lens- and lenslet-prescription samples, measured before and after AR coating.
The coatings improve transmittance through their alumina substrates by \SI{\sim 30}{\percent} in the \SIlist{150;220}{\giga\hertz} SPT-3G observing bands.
Uncertainties in the FTS data are large in the \SI{95}{\giga\hertz} band, but simulations and reflectance measurements suggest the coatings perform similarly there.

\begin{table}[htbp]
  \centering
  \caption{\textbf{Inferred material optical properties and $\mathbf{\pm 1\sigma}$ uncertainties at \SI{150}{\giga\hertz}}}
  \begin{threeparttable}
    \begin{tabular}{l
                    S[table-format=1.3(3)]
                    S[table-format=3.1(1)]
                    S[table-format=1.3(1)]
                    S[table-format=3(1)]}
\toprule
         & \multicolumn{2}{c}{\SI{300}{\kelvin}} & \multicolumn{2}{c}{\SI{77}{\kelvin}} \\
\cmidrule(lr){2-3} \cmidrule(lr){4-5}
Material & {$n$} & {$\tan\delta \times 10^{-4}$} & {$n$} & {$\tan\delta \times 10^{-4}$} \\
\midrule
Porex PM-23J    & 1.292 \pm 0.003 & {$< 1$\tnote{a}} & 1.218 \pm 0.003 & {$< 1$}   \\
Zitex G-115     & 1.234 \pm 0.003 & {$< 1$}          & 1.283 \pm 0.003 & {$< 1$}   \\
Rogers RO3035   & 1.679 \pm 0.002 & 43 \pm 16   & 1.653 \pm 0.002 & 213 \pm 12 \\
Rogers RO3006   & 2.249 \pm 0.003 & 97 \pm 14   & 2.172 \pm 0.004 & 230 \pm 24 \\
CoorsTek AD-995 & 3.112 \pm 0.001 & 49 \pm 6    & 3.089 \pm 0.001 &   3 \pm 1 \\

\bottomrule
    \end{tabular}
  \begin{tablenotes}
    \footnotesize
    \item[a] Due to the sensitivity limits of the experiment, we interpret these values to be an upper limit on the loss tangent for the corresponding materials.
             The Zitex loss tangent is in close agreement with published values \cite{benford:2003}; we expect the Porex to have similar properties.
  \end{tablenotes}
  \end{threeparttable}
  \label{table:matprop}
\end{table}

\begin{table}[htbp]
  \centering
  \caption{\textbf{Sample average transmittance and $\mathbf{\pm 1\sigma}$ uncertainties at \SIlist{300;77}{\kelvin}}}
  \begin{threeparttable}
    \begin{tabular}{lll
                    S[table-format=2(2)]
                    S[table-format=2(1)]
                    S[table-format=3.1(1)]}
\toprule
      &   &   & \multicolumn{3}{c}{Observing band transmittance [\si{\percent}]} \\
\cmidrule{4-6}
T [\si{\kelvin}] & Substrate & State & {\SI{95}{\giga\hertz}\tnote{a}} & {\SI{150}{\giga\hertz}} & {\SI{220}{\giga\hertz}} \\
\cmidrule{1-6}
\multirow{4}*{300} & \multirow[t]{2}*{Lens}                & coated &  90 \pm 70    & 97 \pm 2 & 86.9 \pm 0.4  \\
                   &                                       & bare   &  {$\leq$ 100} & 61 \pm 2 & 57.1 \pm 0.2  \\
                   & \multirow[t]{2}*{Lenslet}             & coated &  {$\leq$ 100} & 94 \pm 3 & 84.6 \pm 0.3  \\
                   &                                       & bare   &  {$\leq$ 100} & 61 \pm 2 & 57.2 \pm 0.4  \\
\midrule
\multirow{4}*{77}  & \multirow[t]{2}*{Lens}                & coated          &  90 \pm 70     & 96 \pm 3      & 85.4 \pm 0.3  \\
                   &                                       & bare            &  {$\leq$ 100}  & 67 \pm 3      & 66.1 \pm 0.5  \\
                   & \multirow[t]{2}*{Lenslet}             & coated\tnote{b} &  {\textemdash} & {\textemdash} & {\textemdash} \\
                   &                                       & bare            &  {$\leq$ 100}  & 67 \pm 3      & 65.0 \pm 0.5  \\

\bottomrule
    \end{tabular}
  \begin{tablenotes}
    \footnotesize
    \item[a] We include the \SI{95}{\giga\hertz} column for completeness, but consider the results to be an upper limit on sample transmittance.
             Uncertainties in these data are large due to the bolometer's responsivity cut-off at \SI{\sim 130}{\giga\hertz}.
    \item[b] The lenslet-prescription coating delaminated from the substrate upon cooldown, so we are missing \SI{77}{\kelvin} data for that sample.
  \end{tablenotes}
  \end{threeparttable}
  \label{table:coateff1}
\end{table}

\begin{figure}[h!]
\centering
\includegraphics[width=8.4cm]{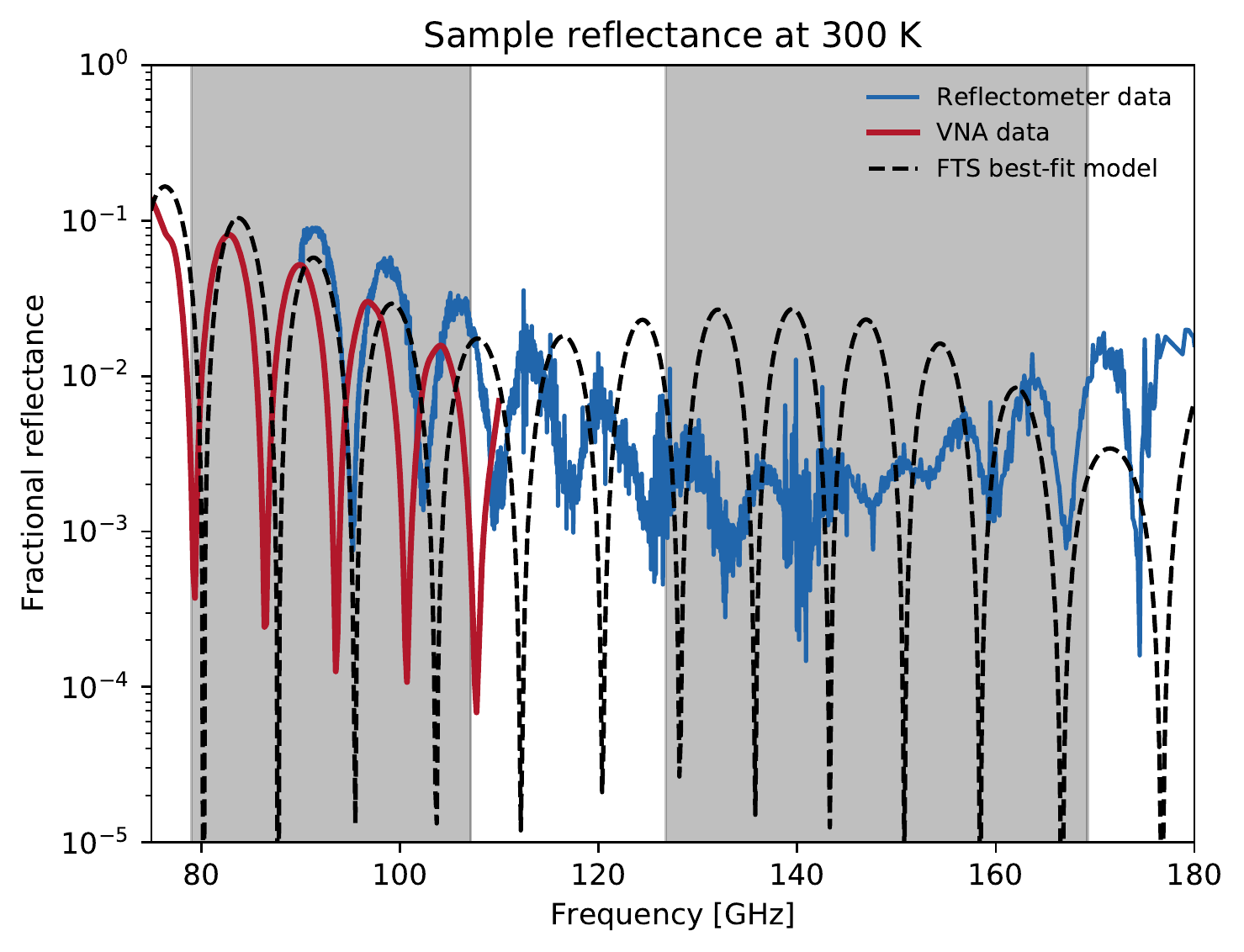}
\caption{
Measured reflectance of lens-prescription coatings at room temperature.
The dashed black line is the best-fit model to our FTS measurements of the lens-prescription coating, the blue curve shows reflectometer data, and the red curve shows single-port VNA data.
The vertical grey bars mark the SPT-3G \SIlist{95;150}{\giga\hertz} observing bands.
The sample substrate used in the FTS and VNA measurements was a different formulation than the substrate used in the reflectometer measurement.
Only accounting for the \SI{10}{\degree} incident angle used in the reflectometer measurement (vs. the normal-incidence VNA measurement) does not explain the difference in the data.
However, accounting for incident angle and tuning the refractive indices of the sample substrates does.
We found that increasing the alumina substrates' refractive index by about \SI{3}{\percent} recovers more of the signal.
But this is a somewhat arbitrary approach and doesn't tell us much, since we should also expect the refractive indices of the AR coating materials to be changing.
}
\label{fig:coatingreflectance}
\end{figure}

We obtained reflectance measurements of the lens-prescription coating in addition to the FTS measurements (\autoref{fig:coatingreflectance}), and present those data here.
We discuss these data in particular in \autoref{subsec:reflectancediscuss}.

\section{Discussion}
\label{sec:discuss}
Measuring the dielectric loss of these materials is challenging, especially if the material under test has a particularly small expected loss tangent, or if it is thin relative to the target wavelength.
Many of the materials are only readily available as thin sheets, so it was necessary to mount them to a substrate before optical testing.
Unfortunately, substrates and bonding layers add additional complexity and degeneracy to the problem.
In this work, we first measured the bare alumina substrates in an attempt to break that degeneracy, but were not completely successful in doing so.
We recognize that the error bars in \autoref{fig:coatingtransmittance} and \autoref{fig:coatingtransmittancecold} are large and that our results would benefit from additional measurements, but such measurements are currently out of our reach for logistical and financial reasons.
Rather than indefinitely delay, we choose to report the results we have now with the hope that other researchers may benefit from our experience.

The lenslet-prescription coating did not survive the cryogenic optical test---the PTFE coating delaminated from one side of the alumina disc on cooldown to \SI{77}{\kelvin}.
Delamination of the lenslet-prescription coating was consistent with our previous experience in cryogenic tests.
We have found the lenslet-prescription coating is robust from \SIrange{\sim 5}{\sim 25.4}{\milli\meter} diameter, but tends to delaminate on cooldown to \SI{77}{\kelvin} at larger diameters (the sample coating used in this work was \SI{\sim 150}{\milli\meter} diameter).
Cutting a perimeter around each \SI{5}{\milli\meter} diameter lenslet allows us to manage the differential thermal contraction between the alumina lenslets, the silicon seating wafer, and the PTFE-based AR coating materials.
In comparison, the lens-prescription coating is robust up to \SI{\sim 700}{\milli\meter} diameter.
We do not have a good physical model for why this is the case, but one hypothesis is that the intermediate LDPE layers of the lens-prescription coating distribute thermal stresses more uniformly throughout the coating.

That the lenslet-prescription coating delaminates at all raises the question of how often the event occurs, particularly at the size scale of the fielded lenslet arrays.
During the 2018/2019 austral summer season we inspected lenslet arrays that had been installed at the SPT-3G survey camera commissioning (austral 2016/2017 summer).
Most of the inspected lenslet arrays had been thermally cycled from \SI{300}{\kelvin} to \SI{\sim 300}{\milli\kelvin} five or more times (accounting for laboratory testing as well as integrated testing at the South Pole).
At that time we found \SI{97.9}{\percent} of the pixels had retained their coatings.
Nearly all cases of coating delamination shared the same cause: the coated pixels were not fully separated from their neighbors at the corners during the laser dicing step.
This was caused by an error in the laser program that we have identified and corrected, so we do not expect this problem going forward.

Though we report a single number for the inferred refractive index ($n$) and loss tangent ($\tan\delta$) in \autoref{table:matprop}, it should be remembered that $n$ and $\tan\delta$ are frequency-dependent quantities.
We take the properties of these materials at \SI{150}{\giga\hertz} as a convenient proxy for their response across the full SPT-3G observing bandwidth (\SIrange{\sim 77}{\sim 252}{\giga\hertz}).
The quantities $n$ and $\tan\delta$ are also temperature-dependent, a distinction we make here.
It is worth noting that while $\tan\delta$ generally decreases with decreasing temperature (for a given frequency and sufficiently wide range of temperature $T$), it does not necessarily do so monotonically.
The functional form of $\tan\delta(T)$ can vary between materials, and the magnitude of $\tan\delta(T)$ may vary between materials that are nominally the same.

Our findings for Zitex G-115's refractive index and loss tangent generally agree with the values reported in \cite{benford:2003}, but we find a lower refractive index and higher loss tangent in our alumina sample than reported in \cite{lamb:1996}.
We expect scattering loss at the porous Zitex/Porex layer to be negligible---even at the highest SPT-3G observing frequencies---due to the small size of the pores (typically \SIrange{1}{3}{\micro\meter}) \cite{zitex_datasheet,priv_comm_porex}.
To the authors' knowledge, measurements of RO3035 and RO3006 at these frequencies have not yet been published.

The increased RO3035 and RO3006 loss at \SI{77}{\kelvin} relative to \SI{300}{\kelvin} is a surprising feature of \autoref{table:matprop} (\num{43e-4} vs. \num{213e-4} and \num{97e-4} vs. \num{230e-4}, respectively).
From experience we have found PTFE, especially loaded PTFE like the Rogers materials, to behave in manners counter to our expectations.
The increased \SI{150}{\giga\hertz} loss we see at \SI{77}{\kelvin} is supported by preliminary measurements of a \SI{\sim 6.35}{\milli\meter} thick sample of RO3006.
We are still working to interpret those data, but the slope of $\tan\delta(T)$ agrees with our findings in this paper.

It is better to avoid multi-material samples and measure a target material in isolation, if the goal is to understand that material's optical properties.
There is less margin for error as optical design tolerances become more stringent, and familiar approximations may not be as useful as we start to use old materials in new ways.
Developing a better understanding of the temperature- and frequency-dependent optical properties of materials (in particular those with very low loss) will require us to try new characterization techniques.
Quasi-optically coupled VNA techniques have been employed at cryogenic temperatures and \si{\tera\hertz} frequencies with good results \cite{zhou:2019}, and would be a good option for us going forward.
In addition, shielded dielectric resonator (SDR) methods could allow us to begin to probe these optical properties, even for thin samples.
SDR methods are an established technique in materials science and well-suited to characterizing low-loss materials \cite{krupka:2006}, but have yet to become standard in the CMB field.

\subsection{Reflectance measurements}
\label{subsec:reflectancediscuss}
The mean reflectance of the lens-prescription sample measured via reflectometer is \SI{\sim 3.1}{\percent} in the probed portion of the \SI{95}{\giga\hertz} SPT-3G observing band, and \SI{< 1}{\percent} across the \SI{150}{\giga\hertz} band.
In comparison, the mean reflectance of the lens-prescription sample measured via VNA is \SI{\sim 2.8}{\percent} throughout the \SI{95}{\giga\hertz} band.
We did not have the necessary equipment (e.g. a frequency extender) to make a VNA measurement in the \SIlist{150}{\giga\hertz} band, so we do not have comparable data there.

As stated in \autoref{subsec:reflectanceexperiment}, the lens-prescription sample measured at the Univerity of Michigan employed a different alumina substrate to the sample measured via FTS and VNA single-port S11.
The substrate used in the reflectometer measurement was not produced by CoorsTek.
We did not measure that specific substrate before coating it, but we have found substrates from the same batch have $n_{\SI{300}{\kelvin}}\approx3.05$, in contrast to the CoorsTek samples which have $n_{\SI{300}{\kelvin}}\approx3.11$.
Therefore, we do not expect the VNA and reflectometer data to match perfectly.

The best-fit optical properties to the FTS data are not well-matched to the S11 data in terms of fringe periodicity; amplitude is a closer match, though still imperfect (\autoref{fig:coatingreflectance}).
Our fit parameters are derived from a higher frequency range than the S11 measurement, so some drift in the optical properties of the materials is anticipated due to the frequency-dependent nature of the dielectric function.

We are able to better model the VNA and reflectometer data by assuming the inferred optical properties of the materials in \autoref{table:matprop} and tuning the refractive indices of the samples' alumina substrates.
In both cases, increasing the substrate refractive indices by \SI{\sim 3}{\percent} almost entirely recovers the fringing periodicity, as well as much of the amplitude.
However, the refractive indices of the other materials should also be changing with frequency, so varying only the alumina refractive index does not tell us much except that the ``effective index'' of each sample appears to increase toward lower frequency.
Carefully quantifying the frequency-dependent dielectric functions describing the optical properties of these materials is beyond the scope of the current analysis and will require further experimentation.

\subsection{Modeling the response of an isolated, AR coated SPT-3G lens and lenslet}
\label{subsec:lensestimate}
For logistical reasons, we did not directly measure the optical efficiency of the finished, AR coated lenses and lenslets in isolation before their integration with the rest of the SPT-3G survey camera.
\autoref{table:model3gcombined} enumerates an effort to use the inferred material properties of the AR coating materials and alumina at \SI{77}{\kelvin} (\autoref{table:matprop}), as well as the known geometry of the SPT-3G lenses and lenslets, to estimate the optical efficiency of those elements after the fact.
We assume a \SI{2.5}{\milli\meter} thick lenslet and \SI{69}{\milli\meter} thick lens for the respective models.
We chose \SI{69}{\milli\meter} because that is the approximate distance traveled by the central ray through the thickest SPT-3G lens (the collimator), which makes the values in \autoref{table:model3gcombined} a conservative performance estimate.
We modeled the optics as multilayer, lossy dielectric slabs using the characteristic matrix method described in \cite{heavens:1955}.
The transmittance (T), reflectance (R), and loss (L) values are the result of \num{10000} realizations of normally incident waves on both lens and lenslet models.
At every realization, $n$ and $\tan\delta$ for each material were randomly sampled from within the $\pm 1\sigma$ regions in \autoref{table:matprop}.
Scattering is not included in the model.
It is important to note that \autoref{table:model3gcombined} does not represent the end-to-end optical efficiency of the SPT-3G experiment, which includes many additional factors.
Rather it suggests the level of performance a CMB experiment with a similar AR coating and lens material might expect from a lens or lenslet.
\begin{table}[htbp]
  \centering
  \caption{\textbf{Modeled response of an SPT-3G lens and lenslet with associated $\mathbf{\pm 1\sigma}$ uncertainty for each SPT-3G observing band}}
  \begin{threeparttable}
    \begin{tabular}{l
                    S[table-format=1.3(3)]
                    S[table-format=1.3(3)]
                    S[table-format=1.3(3)]
                    S[table-format=1.3(3)]
                    S[table-format=1.3(3)]
                    S[table-format=1.3(3)]}
\toprule
  & \multicolumn{3}{c}{2.5 mm-thick lenslet} & \multicolumn{3}{c}{69 mm-thick lens} \\
\cmidrule(lr){2-4} \cmidrule(lr){5-7}
Qty\tnote{a} & \SI{95}{\giga\hertz} & \SI{150}{\giga\hertz} & \SI{220}{\giga\hertz} & \SI{95}{\giga\hertz} & \SI{150}{\giga\hertz} & \SI{220}{\giga\hertz} \\
\midrule
T & 0.955 \pm 0.002 & 0.904 \pm 0.003 & 0.885 \pm 0.004 & 0.677 \pm 0.037 & 0.604 \pm 0.042 & 0.520 \pm 0.045 \\
R & 0.009 \pm 0.002 & 0.045 \pm 0.001 & 0.043 \pm 0.002 & 0.022 \pm 0.001 & 0.024 \pm 0.001 & 0.040 \pm 0.002 \\
L & 0.037 \pm 0.003 & 0.051 \pm 0.003 & 0.072 \pm 0.004 & 0.304 \pm 0.037 & 0.375 \pm 0.042 & 0.444 \pm 0.045 \\
\bottomrule
    \end{tabular}
  \begin{tablenotes}
    \footnotesize
    \item[a] T, R, L refer to transmittance, reflectance, and loss, respectively.
  \end{tablenotes}
  \end{threeparttable}
  \label{table:model3gcombined}
\end{table}

Alumina is the dominant material in both the lens and lenslet systems: it is the thickest material and has the highest refractive index.
The dielectric loss of a sintered alumina part is known to depend on the conditions under which it was manufactured, but there is not a good model to predict a part's loss tangent \textit{a priori} (nor even after the fact in the absence of non-destructive forensic methods).
Since the sintering process itself depends on the geometry of the part, there is no guarantee that the internal microstructure (and by extension, the loss tangent) of a \SIlist{2.5;6.35;25.4;69}{\milli\meter} thick part is identical.
If the loss tangent of an alumina part must be known with high accuracy, then the part should be measured directly---the loss tangent of a proxy such as a witness blank is not necessarily a good indicator, even if it came from the same raw powder batch and firing.

\section{Conclusion}
\label{sec:conclude}
We have developed two prescriptions to produce PTFE-based, broadband, mm-wave AR coatings for cryogenic polycrystalline alumina optics: one for large-diameter optics, the other for arrays of lenslets.
Both coatings increase transmittance through an alumina substrate by \SI{\sim 30}{\percent} in the \SIlist{150;220}{\giga\hertz} SPT-3G observing bands, and are expected to perform similarly in the \SI{95}{\giga\hertz} band.
The total cost of materials needed to AR coat both sides of a \SI{\sim700}{\milli\meter} diameter alumina optic using the method described in this paper is about \$1500 USD.
(This figure excludes the cost of the optic itself.)
The cost of materials required for the lenslet array and lenslet-prescription coating is approximately \$3 USD per pixel ($\sim$\$1 for the coating material, and $\sim$\$2 for the lenslet).

The lamination process developed for lenslets is general to PTFE-based materials, and can be adapted to create coatings for different frequency bands.
A major advantage of the lenslet fabrication process is its throughput.
Using this method, we can fabricate a lenslet array in \SI{\sim 24}{\hour}, which will be important for the future large-scale projects.
While epoxy-based lenslet coatings have also been demonstrated with good results \cite{siritanasak:2018}, fabrication of a single two-layer AR coated lenslet array takes significantly longer and the two-layer coating is not well-suited to experiments with three-band detectors.
The process developed for large-format lenses can also be adapted for different frequencies.
However, the geometry of large-format lenses is a critical factor in the application of the technique described in this paper.
While we have demonstrated the technique in the case of plano-convex lenses, geometries such as meniscus lenses will require additional engineering considerations.

\section*{Funding}
National Science Foundation (NSF) (DGE-1144245, OPP-1248097, PHY-0114422, PHY-1125897); Gordon and Betty Moore Foundation (GBMF 947); U.~S. Department of Energy (DE-SC-0015640)

\section*{Acknowledgments}
J.~V. acknowledges support from an A.~P.~Sloan Foundation fellowship.

\section*{Disclosures}
The authors declare no conflicts of interest.

\theendnotes

\bibliography{arbib}

\begin{thebibliography}{10}
\newcommand{\enquote}[1]{``#1''}

\bibitem{thornton:2016}
R.~J. {Thornton}, P.~A.~R. {Ade}, S.~{Aiola}, F.~E. {Angil{\`e}}, M.~{Amiri},
  J.~A. {Beall}, D.~T. {Becker}, H.~M. {Cho}, S.~K. {Choi}, P.~{Corlies}, K.~P.
  {Coughlin}, R.~{Datta}, M.~J. {Devlin}, S.~R. {Dicker}, R.~{D{\"u}nner},
  J.~W. {Fowler}, A.~E. {Fox}, P.~A. {Gallardo}, J.~{Gao}, E.~{Grace},
  M.~{Halpern}, M.~{Hasselfield}, S.~W. {Henderson}, G.~C. {Hilton}, A.~D.
  {Hincks}, S.~P. {Ho}, J.~{Hubmayr}, K.~D. {Irwin}, J.~{Klein}, B.~{Koopman},
  D.~{Li}, T.~{Louis}, M.~{Lungu}, L.~{Maurin}, J.~{McMahon}, C.~D. {Munson},
  S.~{Naess}, F.~{Nati}, L.~{Newburgh}, J.~{Nibarger}, M.~D. {Niemack},
  P.~{Niraula}, M.~R. {Nolta}, L.~A. {Page}, C.~G. {Pappas}, A.~{Schillaci},
  B.~L. {Schmitt}, N.~{Sehgal}, J.~L. {Sievers}, S.~M. {Simon}, S.~T. {Staggs},
  C.~{Tucker}, M.~{Uehara}, J.~{van Lanen}, J.~T. {Ward}, and E.~J. {Wollack},
  \enquote{{The Atacama Cosmology Telescope: The Polarization-sensitive ACTPol
  Instrument},} {\protect\JournalTitle{Astrophys. J. S.}} \textbf{227}, 21
  (2016).

\bibitem{inoue:2016}
Y.~{Inoue}, P.~{Ade}, Y.~{Akiba}, C.~{Aleman}, K.~{Arnold}, C.~{Baccigalupi},
  B.~{Barch}, D.~{Barron}, A.~{Bender}, D.~{Boettger}, J.~{Borrill},
  S.~{Chapman}, Y.~{Chinone}, A.~{Cukierman}, T.~{de Haan}, M.~A. {Dobbs},
  A.~{Ducout}, R.~{D{\"u}nner}, T.~{Elleflot}, J.~{Errard}, G.~{Fabbian},
  S.~{Feeney}, C.~{Feng}, G.~{Fuller}, A.~J. {Gilbert}, N.~{Goeckner-Wald},
  J.~{Groh}, G.~{Hall}, N.~{Halverson}, T.~{Hamada}, M.~{Hasegawa},
  K.~{Hattori}, M.~{Hazumi}, C.~{Hill}, W.~L. {Holzapfel}, Y.~{Hori},
  L.~{Howe}, F.~{Irie}, G.~{Jaehnig}, A.~{Jaffe}, O.~{Jeong}, N.~{Katayama},
  J.~P. {Kaufman}, K.~{Kazemzadeh}, B.~G. {Keating}, Z.~{Kermish},
  R.~{Keskitalo}, T.~S. {Kisner}, A.~{Kusaka}, M.~{Le Jeune}, A.~T. {Lee},
  D.~{Leon}, E.~V. {Linder}, L.~{Lowry}, F.~{Matsuda}, T.~{Matsumura},
  N.~{Miller}, K.~{Mizukami}, J.~{Montgomery}, M.~{Navaroli}, H.~{Nishino},
  H.~{Paar}, J.~{Peloton}, D.~{Poletti}, G.~{Puglisi}, C.~R. {Raum}, G.~M.
  {Rebeiz}, C.~L. {Reichardt}, P.~L. {Richards}, C.~{Ross}, K.~M. {Rotermund},
  Y.~{Segawa}, B.~D. {Sherwin}, I.~{Shirley}, P.~{Siritanasak}, N.~{Stebor},
  R.~{Stompor}, J.~{Suzuki}, A.~{Suzuki}, O.~{Tajima}, S.~{Takada},
  S.~{Takatori}, G.~P. {Teply}, A.~{Tikhomirov}, T.~{Tomaru}, N.~{Whitehorn},
  A.~{Zahn}, and O.~{Zahn}, \enquote{{POLARBEAR-2: an instrument for CMB
  polarization measurements},} in \emph{\pspie,}  vol. 9914 of \emph{Society of
  Photo-Optical Instrumentation Engineers (SPIE) Conference Series} (2016), p.
  99141I.

\bibitem{hui:2018}
H.~{Hui}, P.~A.~R. {Ade}, Z.~{Ahmed}, R.~W. {Aikin}, K.~D. {Alexander},
  D.~{Barkats}, S.~J. {Benton}, C.~A. {Bischoff}, J.~J. {Bock},
  R.~{Bowens-Rubin}, J.~A. {Brevik}, I.~{Buder}, E.~{Bullock}, V.~{Buza},
  J.~{Connors}, J.~{Cornelison}, B.~P. {Crill}, M.~{Crumrine}, M.~{Dierickx},
  L.~{Duband}, C.~{Dvorkin}, J.~P. {Filippini}, S.~{Fliescher}, J.~{Grayson},
  G.~{Hall}, M.~{Halpern}, S.~{Harrison}, S.~R. {Hildebrandt}, G.~C. {Hilton},
  K.~D. {Irwin}, J.~{Kang}, K.~S. {Karkare}, E.~{Karpel}, J.~P. {Kaufman},
  B.~G. {Keating}, S.~{Kefeli}, S.~A. {Kernasovskiy}, J.~M. {Kovac}, C.~L.
  {Kuo}, K.~{Lau}, N.~A. {Larsen}, E.~M. {Leitch}, M.~{Lueker}, K.~G.
  {Megerian}, L.~{Moncelsi}, T.~{Namikawa}, C.~B. {Netterfield}, H.~T.
  {Nguyen}, R.~{O'Brient}, R.~W. {Ogburn}, S.~{Palladino}, C.~{Pryke},
  B.~{Racine}, S.~{Richter}, R.~{Schwarz}, A.~{Schillaci}, C.~D. {Sheehy},
  A.~{Soliman}, T.~{St. Germaine}, Z.~K. {Staniszewski}, B.~{Steinbach}, R.~V.
  {Sudiwala}, G.~P. {Teply}, K.~L. {Thompson}, J.~E. {Tolan}, C.~{Tucker},
  A.~D. {Turner}, C.~{Umilt{\`a}}, A.~G. {Vieregg}, A.~{Wandui}, A.~C. {Weber},
  D.~V. {Wiebe}, J.~{Willmert}, C.~L. {Wong}, W.~L.~K. {Wu}, E.~{Yang}, K.~W.
  {Yoon}, and C.~{Zhang}, \enquote{{BICEP Array: a multi-frequency degree-scale
  CMB polarimeter},} in \emph{\pspie,}  vol. 10708 of \emph{Society of
  Photo-Optical Instrumentation Engineers (SPIE) Conference Series} (2018), p.
  1070807.

\bibitem{galitzki:2018}
N.~{Galitzki}, A.~{Ali}, K.~S. {Arnold}, P.~C. {Ashton}, J.~E. {Austermann},
  C.~{Baccigalupi}, T.~{Baildon}, D.~{Barron}, J.~A. {Beall}, S.~{Beckman},
  S.~M.~M. {Bruno}, S.~{Bryan}, P.~G. {Calisse}, G.~E. {Chesmore},
  Y.~{Chinone}, S.~K. {Choi}, G.~{Coppi}, K.~D. {Crowley}, K.~T. {Crowley},
  A.~{Cukierman}, M.~J. {Devlin}, S.~{Dicker}, B.~{Dober}, S.~M. {Duff},
  J.~{Dunkley}, G.~{Fabbian}, P.~A. {Gallardo}, M.~{Gerbino},
  N.~{Goeckner-Wald}, J.~E. {Golec}, J.~E. {Gudmundsson}, E.~E. {Healy},
  S.~{Henderson}, C.~A. {Hill}, G.~C. {Hilton}, S.-P.~P. {Ho}, L.~A. {Howe},
  J.~{Hubmayr}, O.~{Jeong}, B.~{Keating}, B.~J. {Koopman}, K.~{Kiuchi},
  A.~{Kusaka}, J.~{Lashner}, A.~T. {Lee}, Y.~{Li}, M.~{Limon}, M.~{Lungu},
  F.~{Matsuda}, P.~D. {Mauskopf}, A.~J. {May}, N.~{McCallum}, J.~{McMahon},
  F.~{Nati}, M.~D. {Niemack}, J.~L. {Orlowski-Scherer}, S.~C. {Parshley},
  L.~{Piccirillo}, M.~{Sathyanarayana Rao}, C.~{Raum}, M.~{Salatino}, J.~S.
  {Seibert}, C.~{Sierra}, M.~{Silva-Feaver}, S.~M. {Simon}, S.~T. {Staggs},
  J.~R. {Stevens}, A.~{Suzuki}, G.~{Teply}, R.~{Thornton}, C.~{Tsai}, J.~N.
  {Ullom}, E.~M. {Vavagiakis}, M.~R. {Vissers}, B.~{Westbrook}, E.~J.
  {Wollack}, Z.~{Xu}, and N.~{Zhu}, \enquote{{The Simons Observatory:
  instrument overview},} in \emph{\pspie,}  vol. 10708 of \emph{Society of
  Photo-Optical Instrumentation Engineers (SPIE) Conference Series} (2018), p.
  1070804.

\bibitem{spt3ginstrumentpaper}
The~South~Pole~Telescope~Collaboration is~preparing a manuscript to~be called, \enquote{{The SPT-3G
  Instrument}}.

\bibitem{abazajian:2016}
K.~N. {Abazajian}, P.~{Adshead}, Z.~{Ahmed}, S.~W. {Allen}, D.~{Alonso}, K.~S.
  {Arnold}, C.~{Baccigalupi}, J.~G. {Bartlett}, N.~{Battaglia}, B.~A. {Benson},
  C.~A. {Bischoff}, J.~{Borrill}, V.~{Buza}, E.~{Calabrese}, R.~{Caldwell},
  J.~E. {Carlstrom}, C.~L. {Chang}, T.~M. {Crawford}, F.-Y. {Cyr-Racine},
  F.~{De Bernardis}, T.~{de Haan}, S.~{di Serego Alighieri}, J.~{Dunkley},
  C.~{Dvorkin}, J.~{Errard}, G.~{Fabbian}, S.~{Feeney}, S.~{Ferraro}, J.~P.
  {Filippini}, R.~{Flauger}, G.~M. {Fuller}, V.~{Gluscevic}, D.~{Green},
  D.~{Grin}, E.~{Grohs}, J.~W. {Henning}, J.~C. {Hill}, R.~{Hlozek},
  G.~{Holder}, W.~{Holzapfel}, W.~{Hu}, K.~M. {Huffenberger}, R.~{Keskitalo},
  L.~{Knox}, A.~{Kosowsky}, J.~{Kovac}, E.~D. {Kovetz}, C.-L. {Kuo},
  A.~{Kusaka}, M.~{Le Jeune}, A.~T. {Lee}, M.~{Lilley}, M.~{Loverde}, M.~S.
  {Madhavacheril}, A.~{Mantz}, D.~J.~E. {Marsh}, J.~{McMahon}, P.~D.
  {Meerburg}, J.~{Meyers}, A.~D. {Miller}, J.~B. {Munoz}, H.~N. {Nguyen}, M.~D.
  {Niemack}, M.~{Peloso}, J.~{Peloton}, L.~{Pogosian}, C.~{Pryke}, M.~{Raveri},
  C.~L. {Reichardt}, G.~{Rocha}, A.~{Rotti}, E.~{Schaan}, M.~M. {Schmittfull},
  D.~{Scott}, N.~{Sehgal}, S.~{Shandera}, B.~D. {Sherwin}, T.~L. {Smith},
  L.~{Sorbo}, G.~D. {Starkman}, K.~T. {Story}, A.~{van Engelen}, J.~D.
  {Vieira}, S.~{Watson}, N.~{Whitehorn}, and W.~L. {Kimmy Wu}, \enquote{{CMB-S4
  Science Book, First Edition},} {\protect\JournalTitle{arXiv e-prints}}
  arXiv:1610.02743 (2016).

\bibitem{abitbol:2017}
M.~H. {Abitbol}, Z.~{Ahmed}, D.~{Barron}, R.~{Basu Thakur}, A.~N. {Bender},
  B.~A. {Benson}, C.~A. {Bischoff}, S.~A. {Bryan}, J.~E. {Carlstrom}, C.~L.
  {Chang}, D.~T. {Chuss}, K.~T. {Crowley}, A.~{Cukierman}, T.~{de Haan},
  M.~{Dobbs}, T.~{Essinger-Hileman}, J.~P. {Filippini}, K.~{Ganga}, J.~E.
  {Gudmundsson}, N.~W. {Halverson}, S.~{Hanany}, S.~W. {Henderson}, C.~A.
  {Hill}, S.-P.~P. {Ho}, J.~{Hubmayr}, K.~{Irwin}, O.~{Jeong}, B.~R. {Johnson},
  S.~A. {Kernasovskiy}, J.~M. {Kovac}, A.~{Kusaka}, A.~T. {Lee}, S.~{Maria},
  P.~{Mauskopf}, J.~J. {McMahon}, L.~{Moncelsi}, A.~W. {Nadolski}, J.~M.
  {Nagy}, M.~D. {Niemack}, R.~C. {O'Brient}, S.~{Padin}, S.~C. {Parshley},
  C.~{Pryke}, N.~A. {Roe}, K.~{Rostem}, J.~{Ruhl}, S.~M. {Simon}, S.~T.
  {Staggs}, A.~{Suzuki}, E.~R. {Switzer}, O.~{Tajima}, K.~L. {Thompson},
  P.~{Timbie}, G.~S. {Tucker}, J.~D. {Vieira}, A.~G. {Vieregg}, B.~{Westbrook},
  E.~J. {Wollack}, K.~W. {Yoon}, K.~S. {Young}, and E.~Y. {Young},
  \enquote{{CMB-S4 Technology Book, First Edition},}
  {\protect\JournalTitle{arXiv e-prints}} arXiv:1706.02464 (2017).

\bibitem{yoon:2008}
K.~W. {Yoon}, \enquote{{Design and deployment of BICEP: a novel small-aperture
  CMB polarimeter to test inflationary cosmology},} Ph.D. thesis, California
  Institute of Technology (2008).

\bibitem{kermish:2012}
Z.~D. {Kermish}, P.~{Ade}, A.~{Anthony}, K.~{Arnold}, D.~{Barron},
  D.~{Boettger}, J.~{Borrill}, S.~{Chapman}, Y.~{Chinone}, M.~A. {Dobbs},
  J.~{Errard}, G.~{Fabbian}, D.~{Flanigan}, G.~{Fuller}, A.~{Ghribi},
  W.~{Grainger}, N.~{Halverson}, M.~{Hasegawa}, K.~{Hattori}, M.~{Hazumi},
  W.~L. {Holzapfel}, J.~{Howard}, P.~{Hyland}, A.~{Jaffe}, B.~{Keating},
  T.~{Kisner}, A.~T. {Lee}, M.~{Le Jeune}, E.~{Linder}, M.~{Lungu},
  F.~{Matsuda}, T.~{Matsumura}, X.~{Meng}, N.~J. {Miller}, H.~{Morii},
  S.~{Moyerman}, M.~J. {Myers}, H.~{Nishino}, H.~{Paar}, E.~{Quealy}, C.~L.
  {Reichardt}, P.~L. {Richards}, C.~{Ross}, A.~{Shimizu}, M.~{Shimon},
  C.~{Shimmin}, M.~{Sholl}, P.~{Siritanasak}, H.~{Spieler}, N.~{Stebor},
  B.~{Steinbach}, R.~{Stompor}, A.~{Suzuki}, T.~{Tomaru}, C.~{Tucker}, and
  O.~{Zahn}, \enquote{{The POLARBEAR experiment},} in \emph{\pspie,}  vol. 8452
  of \emph{Society of Photo-Optical Instrumentation Engineers (SPIE) Conference
  Series} (2012), p. 84521C.

\bibitem{shimura:2017}
F.~{Shimura}, \emph{{Single-Crystal Silicon: Growth and Properties}} (Springer
  International Publishing, Cham, 2017), pp. 1--1.

\bibitem{cavanaugh:2020}
R.~{Cavanaugh}, {Private Communication} (2020).

\bibitem{datta:2013}
R.~{Datta}, C.~D. {Munson}, M.~D. {Niemack}, J.~J. {McMahon}, J.~{Britton},
  E.~J. {Wollack}, J.~{Beall}, M.~J. {Devlin}, J.~{Fowler}, P.~{Gallardo},
  J.~{Hubmayr}, K.~{Irwin}, L.~{Newburgh}, J.~P. {Nibarger}, L.~{Page}, M.~A.
  {Quijada}, B.~L. {Schmitt}, S.~T. {Staggs}, R.~{Thornton}, and L.~{Zhang},
  \enquote{{Large-aperture wide-bandwidth antireflection-coated silicon lenses
  for millimeter wavelengths},} {\protect\JournalTitle{Appl. Opt.}}
  \textbf{52}, 8747 (2013).

\bibitem{mccolm:1990}
I.~J. {McColm}, \emph{{Ceramic Hardness}} (Plenum Press, New York, 1990),
  chap.~6, pp. 247--264.

\bibitem{dobrowolski:2010}
J.~A. Dobrowolski, \emph{{Handbook of Optics, Third Edition Volume IV: Optical
  Properties of Materials, Nonlinear Optics, Quantum Optics (set)}}
  (McGraw-Hill Education, New York, 2010), chap. OPTICAL PROPERTIES OF FILMS
  AND COATINGS.

\bibitem{chamberlin:2012}
R.~A. {Chamberlin} and E.~N. {Grossman}, \enquote{The wintertime south pole
  tropospheric water vapor column: Comparisons of radiosonde and recent
  terahertz radiometry, use of the saturated column as a proxy measurement, and
  inference of decadal trends,} {\protect\JournalTitle{Journal of Geophysical
  Research}} \textbf{117} (2012).

\bibitem{lamb:1996}
J.~W. {Lamb}, \enquote{{Miscellaneous data on materials for millimetre and
  submillimetre optics},} {\protect\JournalTitle{International Journal of
  Infrared and Millimeter Waves}} \textbf{17}, 1997--2034 (1996).

\bibitem{demaine:2009}
E.~D. Demaine, M.~L. Demaine, J.~Iacono, and S.~Langerman, \enquote{Wrapping
  spheres with flat paper,} {\protect\JournalTitle{Computational Geometry}}
  \textbf{42}, 748 -- 757 (2009). Special Issue on the 23rd European Workshop
  on Computational Geometry.

\bibitem{nadolski:2018}
A.~{Nadolski}, A.~M. {Kofman}, J.~D. {Vieira}, P.~A.~R. {Ade}, Z.~{Ahmed},
  A.~J. {Anderson}, J.~S. {Avva}, R.~{Basu Thakur}, A.~N. {Bender}, B.~A.
  {Benson}, J.~E. {Carlstrom}, F.~W. {Carter}, T.~W. {Cecil}, C.~L. {Chang},
  J.~F. {Cliche}, A.~{Cukierman}, T.~{de Haan}, J.~{Ding}, M.~A. {Dobbs},
  D.~{Dutcher}, W.~{Everett}, A.~{Foster}, J.~{Fu}, J.~{Gallichio},
  A.~{Gilbert}, J.~C. {Groh}, S.~T. {Guns}, R.~{Guyser}, N.~W. {Halverson},
  A.~H. {Harke-Hosemann}, N.~L. {Harrington}, J.~W. {Henning}, W.~L.
  {Holzapfel}, N.~{Huang}, K.~D. {Irwin}, O.~B. {Jeong}, M.~{Jonas},
  A.~{Jones}, T.~S. {Khaire}, M.~{Korman}, D.~L. {Kubik}, S.~{Kuhlmann}, C.~L.
  {Kuo}, A.~T. {Lee}, A.~E. {Lowitz}, S.~S. {Meyer}, D.~{Michalik},
  J.~{Montgomery}, T.~{Natoli}, H.~{Nguyen}, G.~I. {Noble}, V.~{Novosad},
  S.~{Padin}, Z.~{Pan}, J.~{Pearson}, C.~M. {Posada}, W.~{Quan}, A.~{Rahlin},
  J.~E. {Ruhl}, J.~T. {Sayre}, E.~{Shirokoff}, G.~{Smecher}, J.~A. {Sobrin},
  A.~A. {Stark}, K.~T. {Story}, A.~{Suzuki}, K.~L. {Thompson}, C.~{Tucker},
  K.~{Vanderlinde}, G.~{Wang}, N.~{Whitehorn}, V.~{Yefremenko}, K.~W. {Yoon},
  and M.~R. {Young}, \enquote{{Broadband anti-reflective coatings for cosmic
  microwave background experiments},} in \emph{\pspie,}  vol. 10708 of
  \emph{Society of Photo-Optical Instrumentation Engineers (SPIE) Conference
  Series} (2018), p. 1070843.

\bibitem{padin:2008}
S.~{Padin}, Z.~{Staniszewski}, R.~{Keisler}, M.~{Joy}, A.~A. {Stark}, P.~A.~R.
  {Ade}, K.~A. {Aird}, B.~A. {Benson}, L.~E. {Bleem}, J.~E. {Carlstrom}, C.~L.
  {Chang}, T.~M. {Crawford}, A.~T. {Crites}, M.~A. {Dobbs}, N.~W. {Halverson},
  S.~{Heimsath}, R.~E. {Hills}, W.~L. {Holzapfel}, C.~{Lawrie}, A.~T. {Lee},
  E.~M. {Leitch}, J.~{Leong}, W.~{Lu}, M.~{Lueker}, J.~J. {McMahon}, S.~S.
  {Meyer}, J.~J. {Mohr}, T.~E. {Montroy}, T.~{Plagge}, C.~{Pryke}, J.~E.
  {Ruhl}, K.~K. {Schaffer}, E.~{Shirokoff}, H.~G. {Spieler}, and J.~D.
  {Vieira}, \enquote{{South Pole Telescope optics},}
  {\protect\JournalTitle{Appl. Opt.}} \textbf{47}, 4418--4428 (2008).

\bibitem{padin:2008b}
S.~{Padin}, J.~E. {Carlstrom}, E.~{Chauvin}, and S.~{Busetti}, \enquote{Panel
  gap cover for millimetre-wave antenna,} {\protect\JournalTitle{Electronics
  Letters}} \textbf{44}, 950--952 (2008).

\bibitem{carlstrom:2011}
J.~E. {Carlstrom}, P.~A.~R. {Ade}, K.~A. {Aird}, B.~A. {Benson}, L.~E. {Bleem},
  S.~{Busetti}, C.~L. {Chang}, E.~{Chauvin}, H.~M. {Cho}, T.~M. {Crawford},
  A.~T. {Crites}, M.~A. {Dobbs}, N.~W. {Halverson}, S.~{Heimsath}, W.~L.
  {Holzapfel}, J.~D. {Hrubes}, M.~{Joy}, R.~{Keisler}, T.~M. {Lanting}, A.~T.
  {Lee}, E.~M. {Leitch}, J.~{Leong}, W.~{Lu}, M.~{Lueker}, D.~{Luong-Van},
  J.~J. {McMahon}, J.~{Mehl}, S.~S. {Meyer}, J.~J. {Mohr}, T.~E. {Montroy},
  S.~{Padin}, T.~{Plagge}, C.~{Pryke}, J.~E. {Ruhl}, K.~K. {Schaffer},
  D.~{Schwan}, E.~{Shirokoff}, H.~G. {Spieler}, Z.~{Staniszewski}, A.~A.
  {Stark}, C.~{Tucker}, K.~{Vanderlinde}, J.~D. {Vieira}, and R.~{Williamson},
  \enquote{{The 10 Meter South Pole Telescope},} {\protect\JournalTitle{{Pub.
  ASP}}} \textbf{123}, 568 (2011).

\bibitem{downey:1984}
P.~M. {Downey}, A.~D. {Jeffries}, S.~S. {Meyer}, R.~{Weiss}, F.~J. {Bachner},
  J.~P. {Donnelly}, W.~T. {Lindley}, R.~W. {Mountain}, and D.~J. {Silversmith},
  \enquote{{Monolithic silicon bolometers},} {\protect\JournalTitle{Appl.
  Opt.}} \textbf{23}, 910--914 (1984).

\bibitem{shoemaker:1980}
D.~H. {Shoemaker}, \emph{A Fourier transform spectrometer for millimeter and
  submillimeter wavelengths} (Massachusetts Institute of Technology, 1980).

\bibitem{mather:1993}
J.~C. {Mather}, D.~J. {Fixsen}, and R.~A. {Shafer}, \emph{{Design for the COBE
  far-infrared absolute spectrophotometer (FIRAS)}} (1993), vol. 2019 of
  \emph{Society of Photo-Optical Instrumentation Engineers (SPIE) Conference
  Series}, pp. 168--179.

\bibitem{chesmore:2018}
G.~E. {Chesmore}, T.~{Mroczkowski}, J.~{McMahon}, S.~{Sutariya}, A.~{Josaitis},
  and L.~{Jensen}, \enquote{{Reflectometry Measurements of the Loss Tangent in
  Silicon at Millimeter Wavelengths},} {\protect\JournalTitle{arXiv e-prints}}
  arXiv:1812.03785 (2018).

\bibitem{barkats:2018}
D.~{Barkats}, M.~I. {Dierickx}, J.~M. {Kovac}, C.~{Pentacoff}, P.~A.~R. {Ade},
  Z.~{Ahmed}, R.~W. {Aikin}, K.~D. {Alexander}, S.~J. {Benton}, C.~A.
  {Bischoff}, J.~J. {Bock}, R.~{Bowens-Rubin}, J.~A. {Brevik}, I.~{Buder},
  E.~{Bullock}, V.~{Buza}, J.~{Connors}, J.~{Cornelison}, B.~P. {Crill},
  M.~{Crumrine}, L.~{Duband}, C.~{Dvorkin}, J.~P. {Filippini}, S.~{Fliescher},
  J.~A. {Grayson}, G.~{Hall}, M.~{Halpern}, S.~A. {Harrison}, S.~R.
  {Hildebrandt}, G.~C. {Hilton}, H.~{Hui}, K.~D. {Irwin}, J.~{Kang}, K.~S.
  {Karkare}, E.~{Karpel}, J.~P. {Kaufman}, B.~G. {Keating}, S.~{Kefeli}, S.~A.
  {Kernasovskiy}, C.~L. {Kuo}, K.~{Lau}, N.~A. {Larsen}, E.~M. {Leitch},
  M.~{Lueker}, K.~G. {Megerian}, L.~{Moncelsi}, T.~{Namikawa}, H.~T. {Nguyen},
  R.~{O'Brient}, R.~W. {Ogburn}, S.~{Palladino}, C.~{Pryke}, B.~{Racine},
  S.~{Richter}, R.~{Schwarz}, A.~{Schillaci}, C.~D. {Sheehy}, A.~{Soliman},
  T.~{St. Germaine}, Z.~K. {Staniszewski}, B.~{Steinbach}, R.~V. {Sudiwala},
  G.~P. {Teply}, K.~L. {Thompson}, J.~E. {Tolan}, C.~{Tucker}, A.~D. {Turner},
  C.~{Umilt{\`a}}, A.~G. {Vieregg}, A.~{Wandui}, A.~C. {Weber}, D.~V. {Wiebe},
  J.~{Willmert}, C.~L. {Wong}, W.~L.~K. {Wu}, H.~{Yang}, K.~W. {Yoon}, and
  C.~{Zhang}, \enquote{{Ultra-thin large-aperture vacuum windows for millimeter
  wavelengths receivers},} in \emph{Proc. SPIE,}  vol. 10708 of \emph{Society
  of Photo-Optical Instrumentation Engineers (SPIE) Conference Series} (2018),
  p. 107082K.

\bibitem{halpern:1986}
M.~{Halpern}, H.~P. {Gush}, E.~{Wishnow}, and V.~{de Cosmo}, \enquote{{Far
  infrared transmission of dielectrics at cryogenic and room temperatures:
  glass, Fluorogold, Eccosorb, Stycast, and various plastics},}
  {\protect\JournalTitle{Appl. Opt.}} \textbf{25}, 565--570 (1986).

\bibitem{heavens:1955}
O.~S. {Heavens}, \emph{Optical Properties of Thin Solid Films}, Dover books on
  physics and mathematical physics (Dover Publications, 1955).

\bibitem{hou:1974}
H.~S. {Hou}, \enquote{Method for optimized design of dielectric multilayer
  filters,} {\protect\JournalTitle{Appl. Opt.}} \textbf{13}, 1863--1866 (1974).

\bibitem{wachtman:1962}
B.~W. J, G.~S. T, and W.~C. G, \enquote{{Linear Thermal Expansion of Aluminum
  Oxide and Thorium Oxide from 100 to 1100K},} {\protect\JournalTitle{Journal
  of the American Ceramic Society}} \textbf{45}, 319--323 (1962).

\bibitem{tonkin:1996}
B.~A. {Tonkin} and M.~W. {Hosking}, \enquote{{The dielectric constant and
  thermal expansion of the ceramic-filled plastic RT Duroid at low
  temperatures},} {\protect\JournalTitle{Journal of Materials Science Letters}}
  \textbf{15}, 2030--2032 (1996).

\bibitem{blumm:2010}
J.~{Blumm}, A.~{Lindemann}, M.~{Meyer}, and C.~{Strasser},
  \enquote{{Characterization of PTFE Using Advanced Thermal Analysis
  Techniques},} {\protect\JournalTitle{International Journal of Thermophysics}}
  \textbf{31}, 1919--1927 (2010).

\bibitem{benford:2003}
D.~J. {Benford}, G.~M. C., and K.~J. W., \enquote{{Optical properties of Zitex
  in the infrared to submillimeter},} {\protect\JournalTitle{Appl. Opt.}}
  \textbf{42}, 5118--5122 (2003).

\bibitem{zitex_datasheet}
Saint-Gobain~Plastics, \enquote{{Microporous PTFE (Zitex G)},}  (2017).
  \url{https://www.films.saint-gobain.com/products/chemfilm/microporous-ptfe}.

\bibitem{priv_comm_porex}
G.~{DiBattista}, {Private Communication} (2015).

\bibitem{zhou:2019}
K.~{Zhou}, S.~{Caroopen}, Y.~{Delorme}, M.~{Batrung}, M.~{Gheudin}, and
  S.~{Shi}, \enquote{Dielectric constant and loss tangent of silicon at
  700–900 ghz at cryogenic temperatures,} {\protect\JournalTitle{IEEE
  Microwave and Wireless Components Letters}} \textbf{29}, 501--503 (2019).

\bibitem{krupka:2006}
J.~Krupka, \enquote{Frequency domain complex permittivity measurements at
  microwave frequencies,} {\protect\JournalTitle{Measurement Science and
  Technology}} \textbf{17}, R55--R70 (2006).

\bibitem{siritanasak:2018}
P.~{Siritanasak}, \enquote{{Precise Measurement of B-mode polarization signal
  from the cosmic microwave background with Polarbear and the Simons Array},}
  Ph.D. thesis, University of California, San Diego (2018).

\end{thebibliography}

\end{document}